\documentclass[num-refs]{wiley-article}
\usepackage{lineno,hyperref}
\pdfoutput=1
\usepackage{graphicx}
\usepackage{amsmath}
\usepackage{amsfonts}
\usepackage{latexsym}
\usepackage{multirow}
\usepackage[algo2e,vlined,linesnumbered]{algorithm2e}

\papertype{FULL PAPER}

\paperfield{Magnetic Resonance in Medicine}

\DeclareCaptionType{suppfigure}[Figure]

\title{\LARGE\textbf{Joint multi-field T\textsubscript{1} quantification for fast field-cycling MRI}}

\author[1,2\authfn{1}]{Markus~B\"odenler}
\author[1\authfn{1}]{Oliver~Maier}
\author[1,5]{Rudolf~Stollberger}
\author[3]{Lionel~M.~Broche}
\author[3]{P.~James~Ross}
\author[4]{Mary-Joan~MacLeod}
\author[1]{Hermann~Scharfetter}
\contrib[\authfn{1}]{Equally contributing authors.}
\corraddress{Markus B\"odenler, Institute of eHealth, University of Applied Sciences FH JOANNEUM, Graz, Austria}
\corremail{markus.boedenler@fh-joanneum.at}

\affil[1]{Institute of Medical Engineering, Graz University of Technology, Graz, Austria}
\affil[2]{Institute of eHealth, University of Applied Sciences FH JOANNEUM, Graz, Austria}
\affil[3]{Aberdeen Biomedical Imaging Centre, University of Aberdeen, Foresterhill, AB25 2ZD, Aberdeen, UK}
\affil[4]{Institute of Medical Sciences, University of Aberdeen, Foresterhill, AB25 2ZD, Aberdeen, UK}
\affil[5]{BioTechMed-Graz, Mozartgasse 12/II, A-8010 Graz, Austria}
\fundinginfo{
ÖAW DOC Fellowship 24966\\
\vspace{5pt}
{\bf WORD COUNT:} Approximately 5000}

\newif\ifisresponse
 \isresponsefalse  
\usepackage{marginnote}
\usepackage{color}
\definecolor{changedcolorEditor}{rgb}{0.3,0.8,0.5}
\definecolor{changedcolorRone}{rgb}{1,0,0}
\definecolor{changedcolorRtwo}{rgb}{0,0.5,0}
\definecolor{changedcolorRthree}{rgb}{0,0,1}
\definecolor{changedcolorRfour}{rgb}{1,0.5,0}
\definecolor{darkolivegreen}{rgb}{0.33, 0.42, 0.18}

\usepackage[draft]{changes}
\ifisresponse

			\newcommand{\responseRtwo}[2]{{\color{changedcolorRtwo}#2}\marginnote{\vspace*{-1.6ex}{\color{changedcolorRtwo}#1}}}

      \newcommand{\mdeleted}[2][None]{\deleted[]{#2}}       

\else

			\newcommand{\responseRtwo}[2]{{\color{black}#2}\marginnote{\vspace*{-1.6ex}{\color{black}}}}

			\newcommand{\mdeleted}[2][None]{\deleted[]{}}	        
\fi

\SetKw{KwBy}{by}
\begin{document}
\setcounter{page}{1}
\maketitle
\begin{abstract}
\textbf{Purpose:} Recent developments in hardware design enable the use of Fast Field-Cycling (FFC) techniques in MRI to exploit the different relaxation rates at very low field strength, achieving novel contrast. The method opens new avenues for in vivo characterisations of pathologies but at the expense of longer acquisition times. To mitigate this we propose a model-based reconstruction method that fully exploits the high information redundancy offered by FFC methods. 
\textbf{Methods:} The proposed model-based approach utilizes joint spatial information from all fields by means of a Frobenius - total generalized variation regulariaztion. The algorithm was tested on brain stroke images, both simulated and acquired from FFC patients scans using an FFC spin echo sequences. The results are compared to three non-linear least squares fits with progressively increasing complexity. 
\textbf{Results:} The proposed method shows excellent abilities to remove noise while maintaining sharp image features with large signal-to-noise ratio gains at low-field images, clearly outperforming the reference approach. Especially patient data shows huge improvements in visual appearance over all fields. 
\textbf{Conclusion:} The proposed reconstruction technique largely improves FFC image quality, further pushing this new technology towards clinical standards.

\end{abstract}


\section{Introduction}
The magnetic field dependency of the longitudinal and transverse relaxation times, also referred to as nuclear magnetic relaxation dispersion (NMRD), provides insight into the underlying structural order and dynamics of a wide range of molecular systems~\cite{Steele2016,Korb2018}. In recent years, the $T_1$ dispersion of protons in particular has experienced increased interest for the investigation of biomarkers related to various pathological processes~\cite{Broche2011,Broche2012,Ruggiero2018,DiGregorio2019}. The field-dependent properties of such biomarkers are invisible to traditional MRI scanners, which operate only at one fixed main magnetic field strength and are restricted to the measurement of relaxation times corresponding to the $B_0$ field employed. However, new MRI-derived technologies are emerging that allow exploring different magnetic fields within a single system. One such technology is Fast Field-Cycling Magnetic Resonance Imaging, also known as FFC-MRI or FFC imaging, which enables a modulation of the main magnetic field during an imaging sequence giving access to field-dependent relaxation properties as a novel contrast mechanism \cite{Lurie2010}. FFC imaging derives from MRI but uses radically different technologies to generate the main magnetic field, and both types of scanner offer different views on biological tissues.

Indeed, varying the main magnetic field within a defined range requires dedicated hardware and various approaches exist to realize FFC imaging systems~\cite{Bodenler2019}. In the clinical field range, FFC imaging is implemented by means of a $B_0$ insert coil together with the superconducting magnet provided by a commercial MRI system for 1.5~T~\cite{Hoelscher2012,Harris2014,Chanet2020} or 3~T~\cite{Bodenler2018}. This approach, also referred to as delta relaxation enhanced MR (dreMR), has auspicious applications for the detection and quantification of contrast agents with increased specificity and sensitivity \cite{Hoelscher2012,Alford2009,Araya2017,Bodenler2019a}. Several systems were also developed to access the endogenous $T_1$ dispersion of tissues in the low-field regime~\cite{Carlson1992,Lurie1998,Ungersma2006,Pine2014,Romero2020}. Recently, a whole-body FFC scanner approved for clinical imaging studies was reported, capable of reaching any field from 50~$\mu$T to 0.2~T~\cite{Broche2019}. Controlled variations of the magnetic field with this single resistive magnet design allow for multi-field $T_1$ quantification over a wide range of field strength while retaining image quality down to ultra-low fields. Pilot studies show promising potential for innovations in the imaging of osteoarthritis~\citep{Broche2012}, sarcoma~\citep{masiewicz2020} or brain stroke~\citep{Broche2019}, with potentially important applications in medicine as an in vivo assessment method of multi-field $T_1$ and $T_1$ dispersion information.

Compared to conventional MRI, implementation of fast field-cycling poses additional demands on power supplies, control electronics, magnet design, pulse sequences and image quality~\cite{Bodenler2019}. Signal-to-Noise Ratio (SNR) is therefore an important issue for FFC scanners to satisfy the latter. High fields benefit from an inherently high SNR as they rely on stable and homogeneous acquisition fields provided by superconducting magnets. Although the SNR is not a limiting factor for the individual images, contrast in dreMR is obtained by image subtraction and strongly depends on the $T_1$ dispersion of the contrast agent in use~\cite{Aime2018}. The magnitude of the dreMR signal is rather small in comparison to the individual images (e.g. about 2.5~\% in~\cite{Bodenler2018}) and retaining sufficient high SNR may become an issue. Similarly, low-field FFC imaging systems operate with acquisition fields of 0.2~T or less, which limits the SNR compared to conventional clinical fields due to its dependency to $B_0$. Moreover, image quality deteriorates because of poor magnetic field homogeneity, field instabilities during operation and delays in the field ramps between different phases in the pulse sequence especially for ultra-low evolution fields. 

For all these reasons, both high- and low-field FFC scanners may strongly benefit from SNR-enhancing methods and a vast number of these have been developed for high field MRI images in the recent years~\citep{coupe2008, ANAND2010842, Fessler2010, Knoll2011, ktslr2011, schloegl2017}. These can be divided into denoising and reconstruction-based approaches. The former takes a series of noisy input images and tries to find a denoised solution by making use of a priori knowledge in the form of regularization. Regularization can utilize either spatial information, information from the acquired series, or a combination of both. Denoising approaches are generally simpler to implement and computation time is lower compared to reconstruction approaches, but they can not recover structures that were missed by the k-space to image space transformation due to poor SNR.
To this end, recent approaches rely on a constrained reconstruction process, incorporating the a priori knowledge in the image generation process~\citep{Fessler2010, Knoll2011, ktslr2011, schloegl2017}. In the case of quantitative MRI, this approach can be taken one step further by including the non-linear MRI signal model into the reconstruction process, thus, directly acting on the parameter maps of interest~\citep{doneva2010, sumpf2011, Wang2017, Maier2019}. This kind of fitting approach is known as model-based reconstruction in the high field MRI regime. Most regularization strategies rely on some sort of sparsifying transform to separate image content from noise and artefacts. Commonly used transforms include finite differences based approaches~\cite{Rudin1992,Bredies2010,Knoll2011} and applications of the wavelet transformation~\cite{Lustig2007,Wang2018,Lai2018}. The regularization functional used highly influences the appearance of the final reconstruction and should be chosen based on a priori knowledge about the given parameter map. It was shown that total generalized variation (TGV)~\citep{Bredies2010} priors are a superb choice for both image reconstruction~\citep{Knoll2011} and quantitative MRI~\cite{Maier2019}, leading to high quality reconstruction results without the stair-casing artefacts of total variation (TV)~\citep{Knoll2011}. Similar to TV, TGV uses information from the image gradient in combination with the assumption that images typically consist of a few, discrete edges and thus fits in the concept of compressed sensing~\cite{Lustig2007}. Opposed to TV, patches between edges are not constrained to have a fixed value but can rather be linearly varying in the case of second order TGV (TGV\textsuperscript{2}). To this end, stair-casing artefacts can be avoided using TGV~\citep{Knoll2011}. Higher order TGV functionals allow for even higher degrees of freedom within the patches but are typically only needed in stereo imaging, such as RGB~\citep{Bredies2020}.

These constrained reconstruction and fitting methods apply well to the estimation of $T_1$ maps. Standard $T_1$ quantification with inversion recovery sequences requires the acquisition of an image series with different inversion times, leading to high redundancy in the information collected that can be exploited by the regularisation algorithm.  FFC imaging adds an extra dimension to MRI by varying the magnetic field during the relaxation phase of the pulse sequence, thus providing an additional field series. This multi-field data offers new possibilities to exploit information redundancies to improve the quantification process. These redundancies could be utilized by a model-based reconstruction approach, incorporating the data from all measurements at different field strengths into one large optimization problem. Each field leads to an individual $T_1$ map which shares common information with the other fields, e.g. most edges in the $T_1$ map should coincide. This information can be exploited by joining the individual regularization functionals in parametric dimension via a Frobenius norm. The Frobenius norm is the matrix equivalent to the $L^2$-norm for vectors and links edge information in parametric dimension. Such an approach was shown to further improve the quality of the resulting parameter maps in the context of $T_1$ mapping from highly subsampled data~\citep{Maier2019}.

Herein, we formulate the multi-field FFC imaging parameter quantification as a non-linear model-based reconstruction problem with Frobenius type TGV\textsuperscript{2} regularization. With this formulation as a single optimization problem it is possible to exploit all the joint spatial information of the additional field dimension to stabilize the quantification process and hence enhance the image quality. The proposed method is evaluated on simulated numerical FFC imaging data as well as on in vivo datasets from two stroke patients and compared to Tikhonov regularized fits from individual fields, all fields combined, and regularization using the squared $L^2$-norm of the gradient (H1-regularization). The results show improved stability of the parameter quantification with excellent noise suppression properties. In particular, the proposed method reveals remarkable contrast between the lesion and surrounding tissues in case of ultra-low fields.  

\section{Theory}

\subsection{Fixing notation}
Throughout the course of this work we fix the 
following notations. The image dimensions in 2D are denoted as $N_x$ and $N_y$, 
defining the image space $U = \mathbb{C}^{N_x \times N_y}$ 
with $p=(x,y)$ defining a point at location $(x,y) \in \mathbb{N}^2$. 
$u \in U^{N_u}$ expresses the space of unknowns $u=\{C, \alpha_{B_0^E} = (\alpha_{B_0^{E_1}}, \alpha_{B_0^{E_2}}, \hdots, \alpha_{B_0^{E_{N_E}}}), T_1^{E} = (T_1^{E_1}, T_1^{E_2}, \hdots, T_1^{E_{N_E}})\} \in U^{N_u}$-maps, 
with $N_u = 1+2N_E$ and $N_E$ the number of evolution fields. The unknowns consist of the scaling $C$, the correction factor for each field $\alpha_{B_0^{E_i}}$, and the field dependent relaxation times $T_1^{E_i}$.
 The measured data space is denoted as $D = \mathbb{C}^{N_{k_x} \times N_{k_y} \times N_d}$ and 
is constituted of $N_d = \sum_{i=1}^{N_E}N_t^{E_i}$ measurements of 2D k-spaces $N_{k_x} \times N_{k_y}$. 
Each measurement is a combination of an evolution field 
$B_0^{E_i} = (B_0^{E_1}, B_0^{E_2}, \hdots, B_0^{E_{N_E}}) \in \mathbb{R}_{+}^{
N_E}$
with an associated number of measurement time points
$t^{evo_i}_n = (t_1^{evo_i}, t_2^{evo_i},\hdots,t_{N_t^{E_i}}^{evo_i}) \in 
\mathbb{R}_{+}^{N_t^{evo_i}}$, and $N_t^{evo_i}$ the number of time points acquired at a specific evolution field $B_0^{E_i}$. 
To simplify notation we will drop the indices and refer to $\alpha_{B_0^E}$ as $\bm{\alpha}$.

\subsection{FFC imaging signal model}
\begin{figure}[ht]
	\centering
	\includegraphics[width=0.95\textwidth]{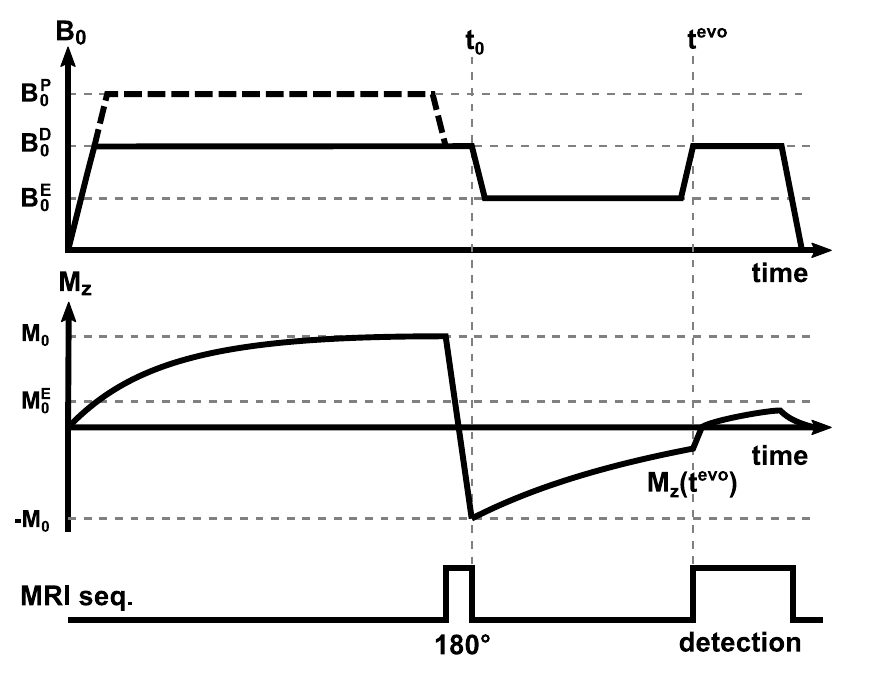} %
	\caption{Pulse sequence diagram of an inversion recovery FFC imaging pulse sequence. A pre-polarization of the sample magnetization can be applied by cycling the main magnetic field to $B_0^P$ (dashed line) or by setting the polarization field to the detection field $B_0^D$, i.e.,~$B_0^P=B_0^D$ (solid line). After an inversion pulse at $t_0$ the longitudinal magnetization $M_z$ evolves at the desired evolution field $B_0^E$ for a given evolution time. The following magnetization $M_z(t^{evo})$ can be detected by any MRI acquisition module. Note that the MRI signal is both inverted and detected at $B_0^D$. $M_0$ and $M_0^E$ represent the equilibrium magnetization for $B_0^D$ and $B_0^E$, respectively.} 
	\label{fig:FFC_seq}
\end{figure}

In the most general case of a FFC imaging pulse sequence, the main magnetic field is rapidly cycled between three different levels: polarization field $B_0^P$, evolution field $B_0^E$ and signal detection field $B_0^D$. A designated pre-polarization of the sample magnetization is not necessarily required in the high SNR regime and the polarization field can be set to the detection field, i.e., $B_0^P=B_0^D$. For simplicity we will assume that this is the case and we will refer to these fields as $B_0$ and the corresponding equilibrium magnetization as $M_0$, respectively, as this does not alter the validity of our approach in the case of low-field systems since the effect of polarisation at a different field can be compensated by a polarisation efficiency term that blends into the inversion efficiency parameter. A schematic of a typical inversion recovery FFC imaging pulse sequence can be seen in Figure \ref{fig:FFC_seq}: following an inversion RF pulse, the main magnetic field is cycled to the desired strength $B_0^E$ and the spin system undergoes a relaxation associated with the applied evolution field during a given evolution time $t^{evo}$. The longitudinal magnetization $M_z$ at the end of this evolution period is given by
\begin{linenomath*}
\begin{align}\label{eq:M_z}
M_z(t^{evo}) = [-\bm{\alpha} M_0-M_0^E]~e^{\frac{-t^{evo}}{T_1^E}} + M_0^E,
\end{align}
\end{linenomath*}
where $M_0$ and $M_0^E$ are the equilibrium magnetizations for the detection and evolution field, respectively. The $T_1$ relaxation time, corresponding to the evolution field applied, is given by $T_1^E$ and $\bm{\alpha}$ corrects $M_0$ for field dependent effects from ramping the field combined with non-ideal inversion efficiency of the RF pulse~\citep{signaleq}. The equilibrium magnetization is proportional to the strength of the applied magnetic field, i.e., $M_0=C\,B_0$ and $M_0^E=C\,B_0^E$ for the detection and evolution field, respectively. This can also be written as      
\begin{linenomath*}
\begin{align}\label{eq:M_0}
\frac{M_0}{B_0}=\frac{M_0^E}{B_0^E}=C\,.
\end{align}
\end{linenomath*}
After the evolution period the signal is acquired at $B_0$ to ensure that the Larmor frequency of the spins corresponds to the tuning frequency of the receive RF coil. With equations (\ref{eq:M_z}) and (\ref{eq:M_0}), and the unknown parameters $u=(C,\bm{\alpha},T_{1}^E)$, the acquired signal $S(u)$ can be modeled for a specific evolution field $B_0^E$ and evolution time $t^{evo}$ by the non-linear signal equation $S: U \to D$, given by

\begin{linenomath*}
\begin{align}
\label{eq:sig}
S(u) = \mathcal{F}\left\{C\,[- \bm{\alpha}\,B_0\,e^{\frac{-t^{evo}}{T_{1}^{E}}} + 
B_0^E\,(1-e^{\frac{-t^{evo}}{T_{1}^E}})]\right\},
\end{align}
\end{linenomath*}
with $\mathcal{F}$ representing the Fourier transformation and sampling of k-space. 

\subsection{Multi-field parameter fitting}
Acquiring several time points $t^{evo}$ for a specific evolution field $B_0^E$ allows 
to quantify $C$, $\bm{\alpha}$, and $T_{1}^E$. Typically, each
$B_0^E$ field yields a different $T_{1}^E$ value, thus the fitting process must be
repeated for each evolution field, omitting joint information in the 
unknowns such as structural information. By combining the separate
fitting steps into a single optimization problem it is possible to utilize 
the shared information to stabilize the quantification process. 
Furthermore, joint information between different parameter maps can
be exploited by means of a Frobenius type functional. Such a fitting and 
regularization strategy has been successfully applied in other multi-channel fitting problems 
~\citep{Bredies2014,Knoll2017,Maier2019}. Thus we propose to apply
a similar approach to quantify $C$, $\bm{\alpha}$, and $T_{1}^E$ from multi-field
FFC imaging data, utilizing shared information, especially between the different 
$T_{1}^E$ maps. 

\subsection{Model-based reconstruction framework}\label{sec:opt}
Assuming Gaussian noise corrupts the measurement data $d$, 
it is possible to quantify the unknown parameter $u$ via a 
regularized non-linear, minimum-least-squares problem
\begin{linenomath*}
\begin{equation}\label{eq:l2minreg}
   \underset{u}{\min}\quad \frac{1}{2}
  \left\|A(u)-
  d\right\|_2^2 + \gamma R(u),
\end{equation}
\end{linenomath*}
which origins from a maximum a posteriori approach using Bayes' theory. $A$ denotes some non-linear forward operator and $R$ reflects a priori knowledge about the unknowns $u$ by means of a regularization term. $\gamma$ can be used to weight between data and regularization term and may either be a scalar value or a vector for each unknown in $u$.

For multi-field FFC data $u$ 
is linked to the measurement data 
$d=(d_{1,1},d_{1,2},\hdots,d_{1,N_t^{E_1}},d_{2,N_t^{E_1}}, \hdots, 
d_{N_E,N_t^{E_{N_E}}})\in U$ via 
$S\responseRtwo{R2.C8}{\mdeleted{_{B^{E_i}_0,t^{E_i}_n}}}: u \mapsto d_{i,n}$
We denote all measured data as $\bm{d}$ and the corresponding mapping from unknown space to data space as $\bm{S}$. Thus, 
 the optimization problem is defined as
 \begin{linenomath*}
\begin{align}
   \underset{u,v}{\min}\quad 
\frac{1}{2}\|
\bm{S}(u)-\bm{d}\|_2^2 + \gamma( \beta_0\|\nabla u - v\|_{1,2,F} + 
\beta_1\|\mathcal{E}v\|_{1,2,F}).
\end{align}
\end{linenomath*}
$R(u)$ is chosen as TGV$^2$ regularization with a joint Frobenius norm
on all unknowns $u$. $\mathcal{E}$ denotes the finite symmetric derivative and the auxiliary variable $v$ balances between the first and second derivative of the TGV$^2$ functional. This type of regularization was shown to have 
favorable properties for multi-parameter model-based 
reconstruction~\citep{Maier2019}.
The $\|\cdot\|_{1,2,F}$ terms resemble the Frobenius type TGV$^2$ functionals, 
joining common spatial information of the unknown parameter maps by combining gradient information of
all maps via an $L^2$-norm in parameter direction.

Using the TGV$^2$ model parameters $\beta_0$ and $\beta_1$ it is possible to 
balance the approximated first and second derivatives, 
avoiding the stair-casing artifacts of TV while maintaining 
its favorable edge-preserving features. The ratio $\beta_0/\beta_1=1/2$ of 
TGV$^2$ is fixed throughout this work, as it was shown to yield good results
for MRI image reconstruction~\citep{Knoll2011}. The numerical solution 
is done in analogy to~\citep{Maier2019} via a 
Gauss-Newton (GN) approach. Leading to an inner GN problem of the form
\begin{linenomath*}
\begin{align} \label{eq:linearized}
   \underset{u,v}{\min}\quad &\|
\bm{DS}u-\tilde{\bm{d}}^k\|_2^2 
+ \nonumber\\
&\gamma_k(
\beta_0\|\nabla u - v\|_{1,2,F} + \beta_1|\|\mathcal{E}v\|_{1,2,F}) +
\nonumber\\ &\frac{\delta_k}{2}\|u-u^k\|_{M_k}^2.
\end{align}
\end{linenomath*}
The linearization is done via a Taylor series expansion of 
$\bm{S}$ w.r.t. each unknown in $u$ at position $u^k$.
Constant terms are fused into $\tilde{\bm{d}}^k$ to keep the notation clean.
$\bm{DS}$ amounts to the Jacobian of $\bm{S}$, evaluated at $u^k$.
Introducing an additional weighted $L^2$-norm penalty on $u$ improves 
convexity of the function. The weighting matrix $M_k$ can be used to 
resemble a Levenberg-Marquart update if $M_k=diag(\bm{DS}^T\bm{DS})$.
The regularization parameters $\gamma_k$ and $\delta_k$ balance between the three terms and are reduced
after each linearization step. Reducing the weights was shown to be beneficial in the context of the IRGN algorithm~\cite{Kaltenbacher2010}.

Using Fenchel duality the problem of non-differentiability can be overcome and
equation~\ref{eq:linearized} can be cast into a 
saddle-point form
\begin{linenomath*}
\begin{equation}\label{eq:PD}
\underset{u}{\min}\,\underset{y}{\max}~ \left<\mathrm{K}u,y\right> + G(u) - 
F^*(y).
\end{equation}
\end{linenomath*}
$K$ constitutes the linear operators encountered in equation~\ref{eq:linearized} within data and TGV$^2$ norm, $G$ reflects the quadratic penalty on $u$, and $F^*$ denotes the dual norms of the data and TGV$^2$ term.
Problems in such a form can be solved via a primal-dual 
algorithm~\citep{Chambolle2011} using a line-search to speed up 
convergence~\citep{Malitsky2018}. Mathematical details for each step are given in Supporting~Information~Text~A.
The update scheme written as pseudo code is given in Supporting~Information~Text~B.

\section{Methods}
\subsection{Numerical FFC imaging data}
\begin{figure}[ht]
	\centering
	\includegraphics[width=0.95\textwidth]{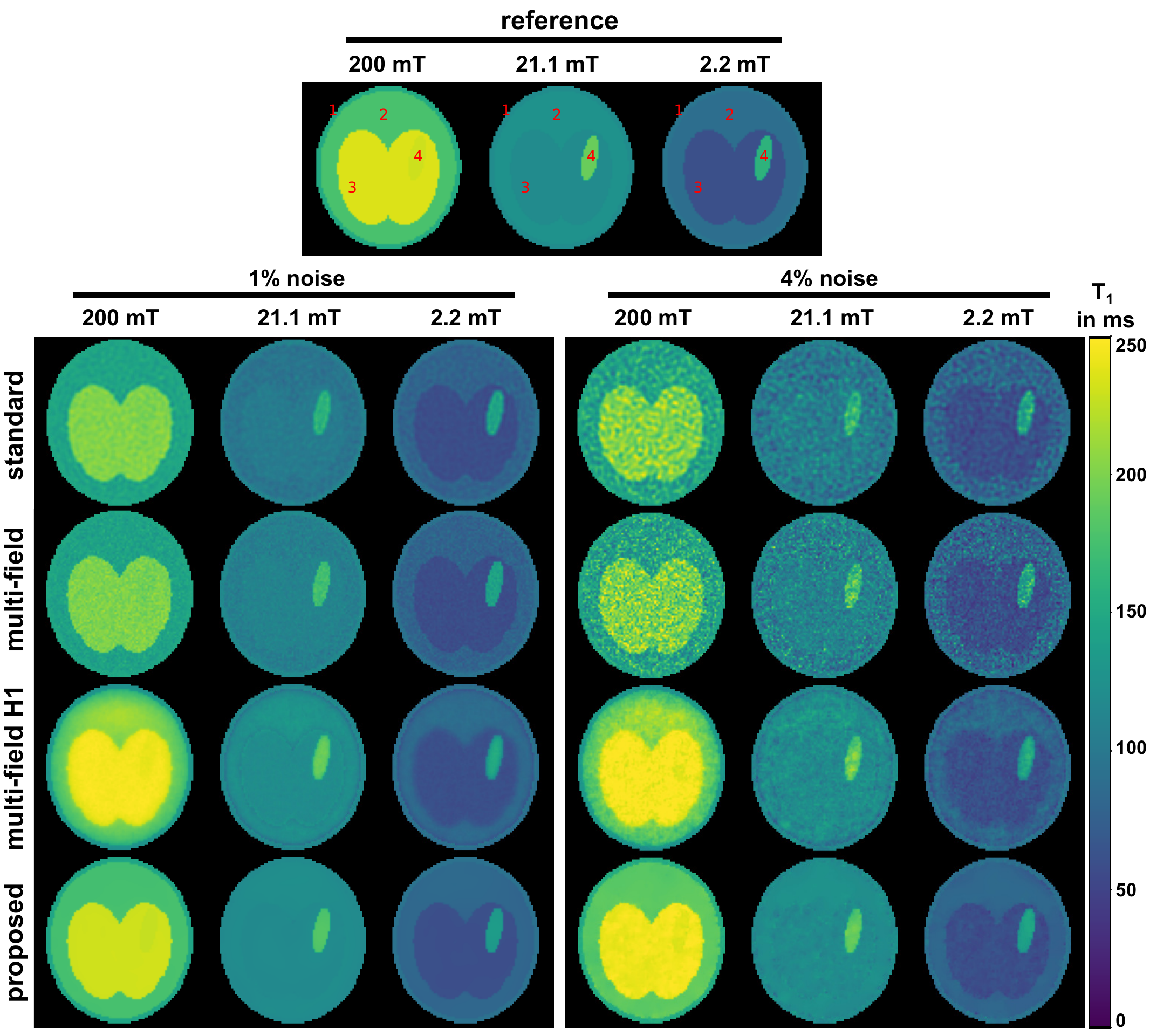} 
	\caption{Multi-field $T_1$ maps obtained from simulated FFC imaging inversion recovery data. The reference $T_1$ maps for three different evolution fields (200~mT, 21.1~mT and 2.2~mT) are shown at the top.
	The different reconstruction methods are presented in each row. Standard refers to single field pixel-wise fitting, multi-field to combined field, pixel-wise fitting approach and H1 to the model-based approach with regularization using the squared $L^2$-norm of the gradient. The proposed method is shown in the last row. The columns show increasing noise from left to right.}	\label{fig:phantom_T1}
\end{figure}

To evaluate the proposed model-based reconstruction approach numerical FFC imaging data were simulated using parameters measured from FFC imaging scans of brain stroke patients at the University of Aberdeen as part of a separate study (PUFFINS study, see details in section~\ref{sec:invivoffc}). The numerical phantom followed a schematic geometry and dispersive characteristics of an axial head scan with four regions (see Figure~\ref{fig:phantom_T1}), representing the subcutaneous fat (region of interest (ROI) 1), the tissues surrounding the brain (ROI 2), the brain (ROI 3) and a stroke-like lesion (ROI 4). $T_1$ values were simulated by means of a power-law dispersion with model parameters $a$ and $b$, $1/T_1=a({B^E_0})^b$, coarsely in line with proton $T_1$-NMRD profiles of fat (ROI 1), white (ROI 2) and grey matter (ROI 3) and stroke lesions (ROI~4) measured in vivo from the PUFFINS patients cohort (data to be published). The values retained for the different evolution fields and times are summarized in Table~\ref{tab:table-simlations} and Table~\ref{tab:table-T1-simulations}, respectively, and served as a ground truth for the validation of the $T_1$ quantification. The numerical FFC imaging phantom was first generated as a vector graphic to be subsequently converted to matrix data to allow for any desired sampling resolution. In this case we used an image resolution with matrix size of $128\times128$ pixel, which is typical for the original FFC imaging of stroke patients. Tissue reference values were assigned for each ROI and a data series was generated using the signal equation in~\ref{eq:sig}. Additionally, a small constant phase offset was introduced for each $\alpha$.

Simulated proton density values were normalized to 1, resulting in a theoretical maximum signal amplitude of 1 for the simulated series.
Zero-mean Gaussian noise was added on both real and imaginary parts of the images to simulate the noise arising from the patient tissues and acquisition system. The noise amplitude was selected as a percentage of the theoretical maximum signal in the overall ground truth image series ranging from 1~\% to 4~\%, reflecting a typical range from the FFC images acquired. SNR directly after inversion at $B_0$ ranged from 33.3 to 8.3 in the white matter ROI and 66.7 to 16.7 for the grey matter ROI, respectively. Note that low-field FFC images tend to exhibit markedly lower signal strength than higher-field ones because of losses when the magnetic field switches, so their SNR was proportionally more affected using this approach. Finally, the image series was transformed to k-space via a 2D Fourier transformation, as input for the proposed method with TGV$^2$ and H1 regularization, respectively. The pixel-wise fitting methods were applied to the image series.

\subsection{In vivo FFC imaging data}
\label{sec:invivoffc}
\begin{figure}[ht]
	\centering
	\includegraphics[width=0.95\textwidth]{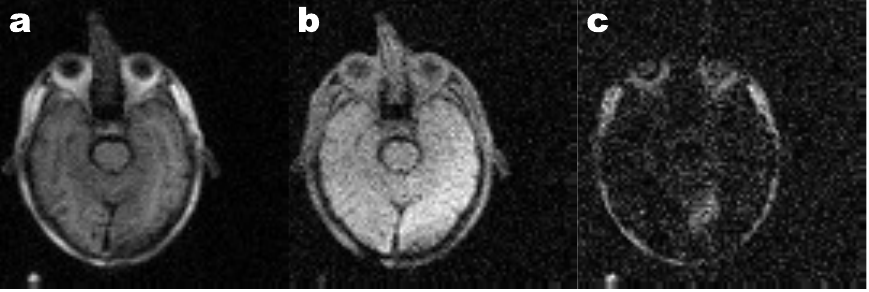}
	\caption{FFC images of a stroke patient from the PUFFINS study (patient I). Image obtained at 200 mT evolution field show good signal after inversion (b) and after 455~ms evolution time (a) but low contrast in the lesion, while low-field images at 21~mT (c) show good lesion contrast but most other tissue shows little to no signal.}
	\label{fig:invivo_data}
\end{figure}

The performance of the proposed method was tested on in vivo FFC imaging patient data. Two data sets obtained from patients scanned for a brain stroke were selected, as part of the PUFFINS study currently taking place at the University of Aberdeen. This study has been approved by the North of Scotland Research Ethics Committee (study number 16/NS/0136) and all the participants agreed for the clinical and FFC imaging data to be used anonymously for research purposes. The scans selected both present a lesion in the ultra-low field regime that could not be easily observed at 200 mT, as illustrated in Figure \ref{fig:invivo_data} for patient I. Both cases were assessed from computed tomography (CT) and diffusion-weighted MRI scans as embolic stroke for patient I and multiple embolic events for patient II. FFC measurements were performed using a whole-body FFC scanner~\citep{Broche2019} using a FFC inversion-recovery spin echo sequence~\citep{Ross2014} with an echo time of 24 ms, 20 kHz bandwidth, 8.37 MHz acquisition frequency, 10 mm slide thickness and single slice acquisition. The images had a field of view of 290 mm and a resolution of 128 x 128 pixel in-plane with 80 phase encode acquisitions and partial Fourier acquisition (80 lines out of 128). The sample was pre-polarised at 200 mT for 300 ms before each evolution periods with the timings as shown in Table ~\ref{tab:table-simlations}, for an acquisition time of 40 min.

\subsection{Data processing and corrections}
The original raw image was reconstructed using partial Fourier completion to recover the correct image ratio. Phase-encode artefacts were removed using a method previously published~\citep{broche2017} but the images had not been filtered or further modified. While the noisy images were used as input for the standard pixel-based fitting, the corresponding noisy k-space data was used as input for the proposed fitting process using H1 and TGV$^2$ regularization, respectively. 

As reference method \textit{lsqnonlin} of Matlab (The MathWorks, Inc.) was used for fitting equation~\ref{eq:l2minreg} field-by-field and pixel-by-pixel, where $R(u)$ was replaced with Tikhonov regularization on the unknowns to stabilize fitting. Prior to fitting, images were smoothed in k-space using the following filter function

\begin{linenomath*}
\begin{equation}
    f(\bm{k}) = \frac{1}{2}+\frac{1}{\pi}\arctan{\beta\frac{k_c-|\bm{k}|}{k_c}}, 
\end{equation}
\end{linenomath*}
with $k_c=30$ denoting the cutoff radius, $\bm{k}$ the k-space location, and $\beta=100$ as parameter for the slope of the filter.

As a second reference method, all fields were combined for the pixel-wise fitting without pre-smoothing, similar to the proposed method. As a third reference, an H1 regularization was used as $R(u)$, i.e. penalizing the squared $L^2$-norm of the gradient of $u$. The latter approach was implemented in Python and optimized using the proposed IRGN algorithm with an accelerated gradient descent optimizer for the inner iterations. Again, no pre-smoothing was applied.

The analyses with the proposed method were done by implementing the FFC signal model in PyQMRI~\citep{Maier2020}. All fittings were performed on a desktop PC equipped with an Intel(R) Core(TM) i7-6700K CPU @ 4.00GHz with 64 gigabyte of RAM and a NVIDIA GeForce GTX 1080 Ti GPU with 12 gigabyte of RAM.

\subsection{Optimization}
The regularization weights $\gamma_k$ and $\delta_k$ were reduced after each linearization step, following the iterative regularized Gauss-Newton scheme~\citep{jin_zhong_2013}.
$\gamma_k=10^{-3}$ and $\delta_k=1$ were used as initial values and were reduced by a factor of $0.5$ and $0.1$, respectively. To account for the typical smooth appearance of $\bm{\alpha}$, corresponding regularization weights were multiplied by a factor of 10. The reduction steps were repeated down to $\gamma_{min}=4\times10^{-6}$ and $\delta_{min}=10^{-3}$. In total, 12 linearization steps were performed. The number of primal-dual iterations for each sub-problem was doubled starting at 10 iterations up to 2000 iterations, i.e. $iter_k = min(10*2^k, 2000)$. If the relative decrease in the primal problem or the decrease of the primal-dual gap was less than $10^{-6}$, the inner iteration was terminated. The step sizes of the employed primal-dual algorithm were determined via a line-search, described in \textit{Algorithm 2}~\citep{Malitsky2018}. The same approach and the same weights have been used for the H1-regularized reference method. Weights for the Tikhonov based approaches have been selected as small as possible to still achieve a stable fitting (${2\cdot10^{-11}}$).

\section{Results}
\subsection{Numerical FFC imaging data}

\begin{figure}[ht]
	\centering
	\includegraphics[width=0.95\textwidth]{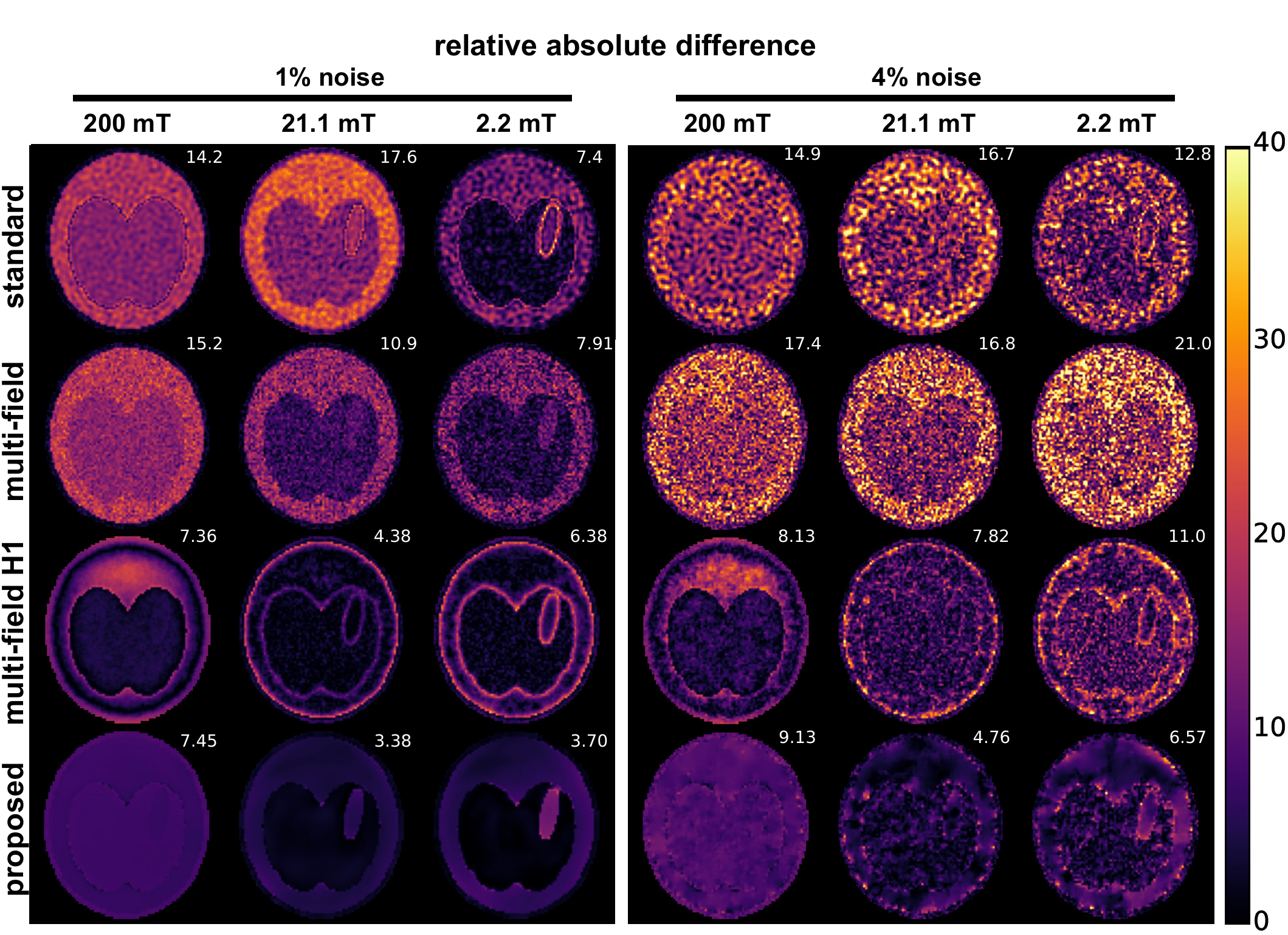} 
	\caption{Pixel-wise relative absolute difference to the ground truth $T_1$ values. Numbers next to the difference images show mean relative absolute error within the phantom. All values are given in percent. }
	\label{fig:phantom_diff}
\end{figure}

\begin{figure}[ht]
	\centering
	\includegraphics[width=0.95\textwidth,height=0.85\textheight, keepaspectratio]{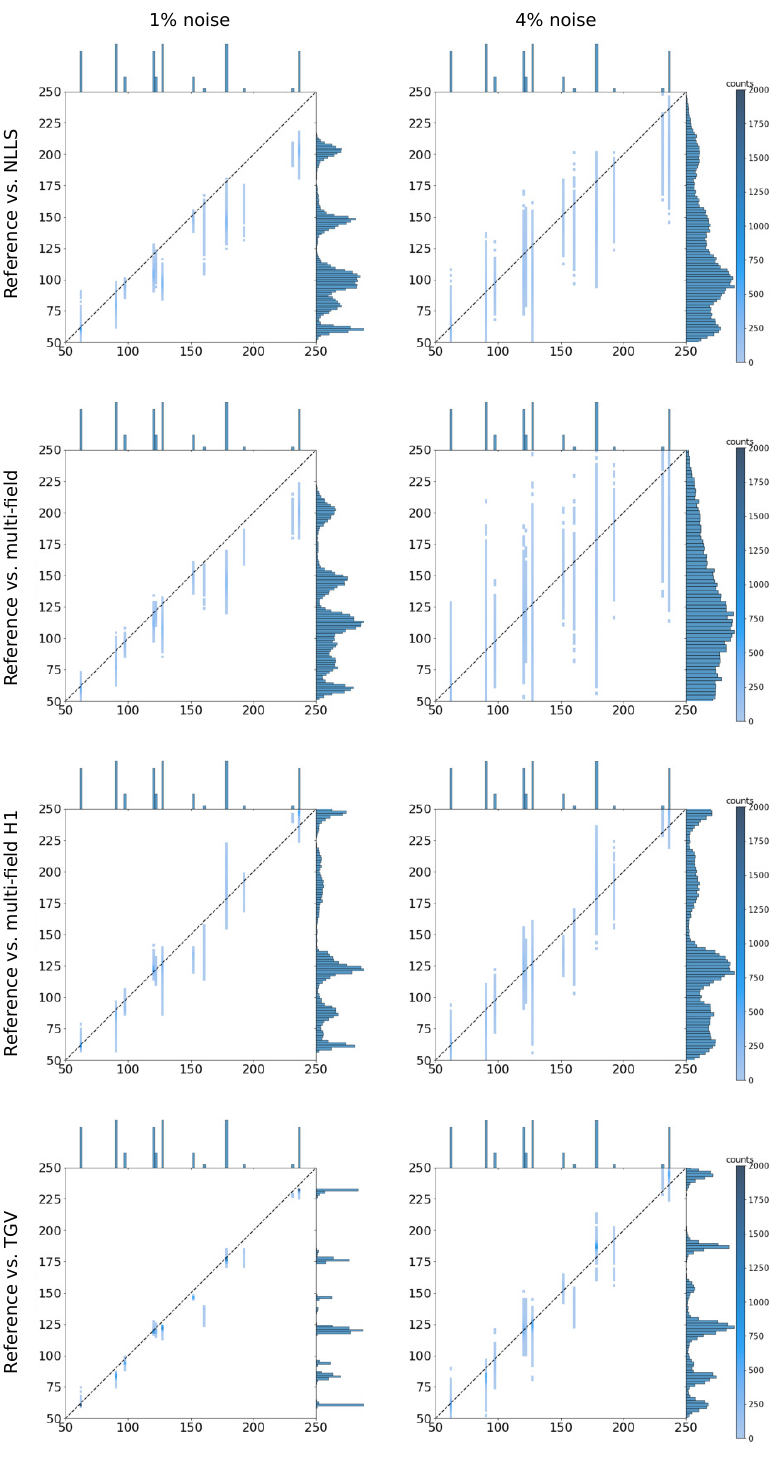}
	\caption{2D histogram evaluation for $T_1$ maps in Figure~\ref{fig:phantom_T1}, which were obtained from synthetic FFC imaging data by pixel-wise fitting (standard), combined field pixel-wise fitting (multi-field), multi-field H1, and joint model-based reconstruction (proposed). The dashed line represents identity. Shown are reference values on the ordinate versus results obtained with the different reconstruction methods on the abscissa. Points below the identity line correspond to under-estimation, points above to over-estimation, respectively. All values are given in ms.}	\label{fig:phantom_T1_2Dhist}
\end{figure}

The simulated high noise level can be seen as residual noise in the reconstructed $T_1$ maps of the pixel-wise fitting approaches (Figure~\ref{fig:phantom_T1}). Simultaneously, a difference to the simulated reference is visually noticeable in the pixel-wise fitting approach. The H1 approach is able to reduce these outliers but suffers from blurring at image edges. The proposed model-based method is able to reduce outliers throughout all noise levels and is visually closer to the simulated reference values. Plots of $C$ and $\bm{\alpha}$ in Supporting~Information~Figures~\ref{fig:phantom_C_abs}-\ref{fig:phantom_alpha_angle} show similar results. The single-field pixel-wise fitting approach even fails to capture the correct phase of the simulated phantom. A pixel-wise relative absolute difference plot (Figure~\ref{fig:phantom_diff}) confirms this visual impression of reduced noise using the proposed approach. The proposed method shows an up to 18 fold lower mean error in the phantom, computed over all pixels, than standard pixel-wise fitting. The error increases with increased noise level, as can be expected. Difference plots also reveal a slight bias of the proposed method. The bias of the methods is further assessed in 2D joint histogram plots (Figure~\ref{fig:phantom_T1_2Dhist}). For these plots, $T_1$ values of all fields are combined to form a single plot. The proposed method shows slight underestimation of high $T_1$ values, as reflected by points lying below the identity line. However, noise could be greatly reduced compared to the pixel-wise fitting and different $T_1$ ranges are clearly separated and show a similar distribution as the simulated values. Fitting with the standard method took approximately 100 seconds. The proposed method took roughly 120 seconds.

\subsection{In vivo FFC imaging data}
\begin{figure}[ht]
	\centering
	\includegraphics[width=0.95\textwidth, height=0.8\textheight, keepaspectratio]{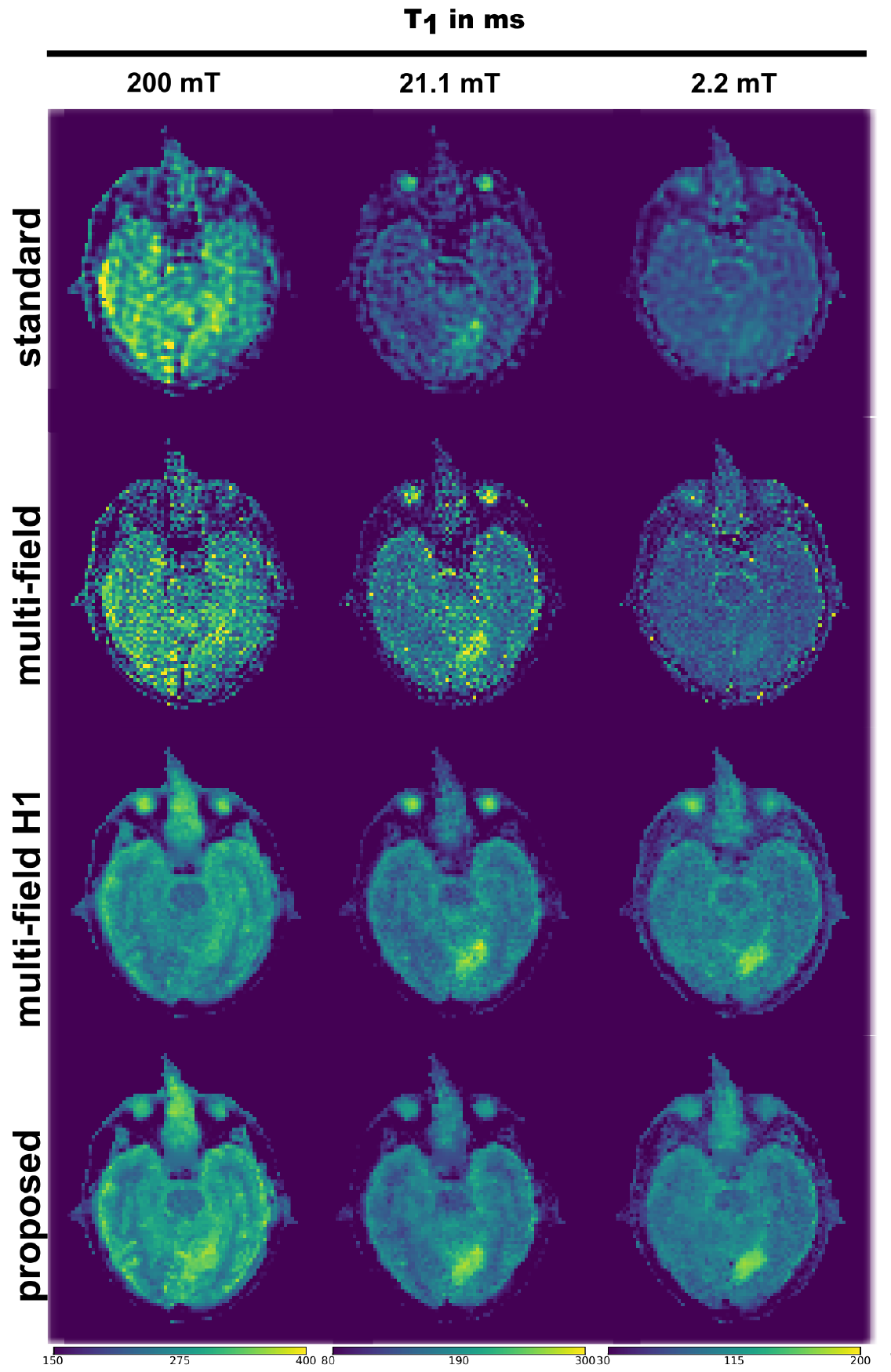}
	\caption{In vivo multi-field $T_1$ maps of a transverse slice of the brain of stroke patient I. From top to bottom, $T_1$ maps were obtained at three different evolution fields $B_0^E=\{200,~21.1,~2.2\}~mT$ by pixel-wise fitting of the signal model for each $B_0^E$ separately, combined field pixel-wise fitting, multi-field model-based reconstruction with H1 regularization and by the proposed multi-field model-based reconstruction approach utilizing the joint information of all three evolution fields (bottom row).}	\label{fig:invivo_T1_dataset1}
\end{figure}

\begin{figure}[ht]
	\centering
	\includegraphics[width=0.95\textwidth,height=0.8\textheight, keepaspectratio]{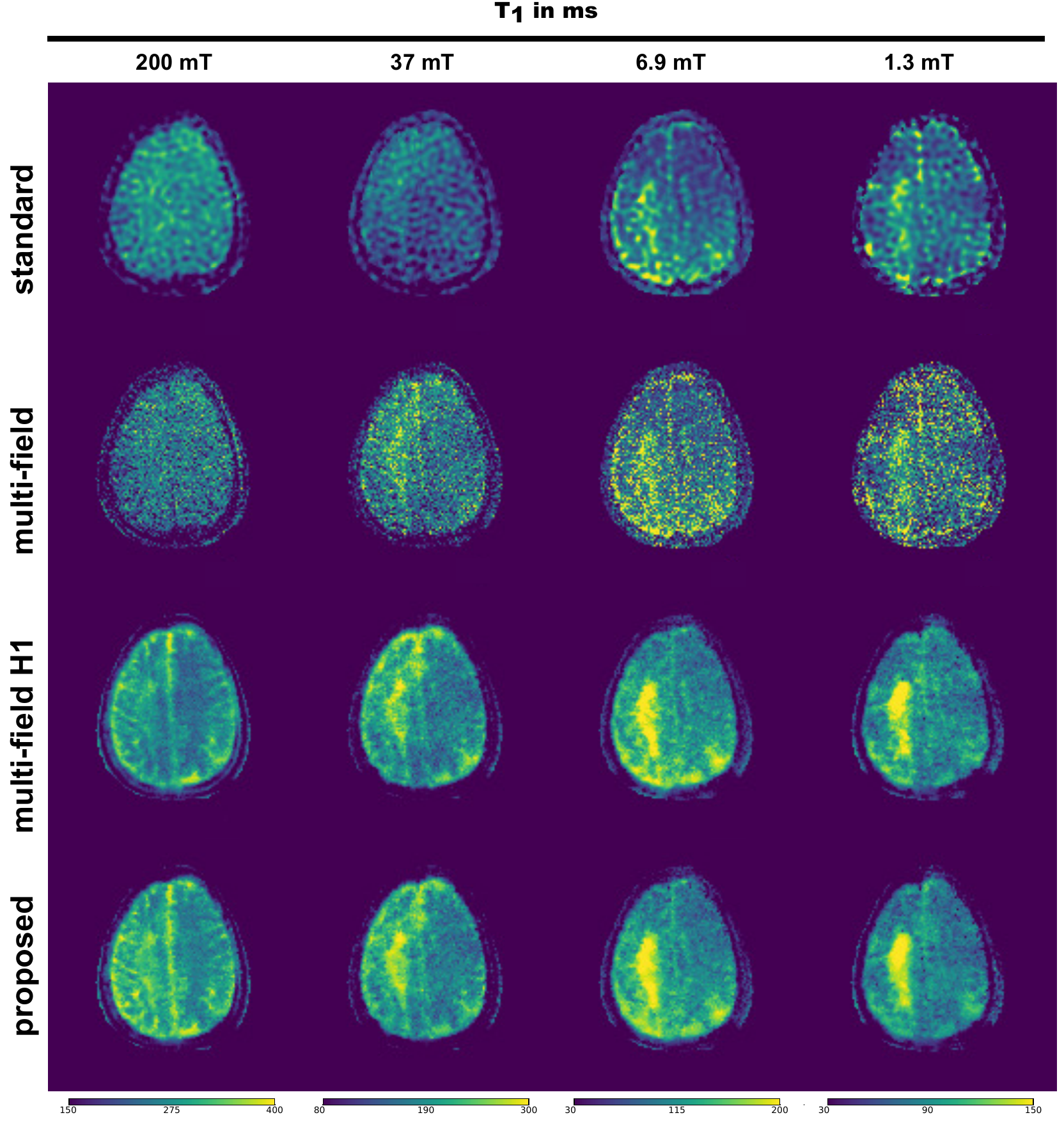} 
	\caption{In vivo multi-field $T_1$ maps of a transverse slice of the brain of stroke patient II. $T_1$ maps were obtained at four different evolution fields $B_0^E=\{200,~37,~6.9,~1.3\}$ mT. The different reconstruction methods are given in each row. From top to bottom the methods are pixel-wise fitting of the signal model for each $B_0^E$ separately, combined field pixel-wise fitting, multi-field model-based reconstruction with H1 regularization and by the proposed multi-field model-based reconstruction approach utilizing the joint information of all three evolution fields.}	\label{fig:invivo_T1_dataset2}
\end{figure}

The improvements in $T_1$ estimations held true when processing real FFC imaging data from stroke patients. The $T_1$ maps of unfiltered FFC images obtained using standard fitting-based processing methods could not resolve anatomical features inside the brain region, as seen in Figures \ref{fig:invivo_T1_dataset1} and \ref{fig:invivo_T1_dataset2}. Spatial regularization in combination with multi-field fitting could greatly improve image contrast. The proposed method offers clear distinguishable structures in $T_1$ maps at 200 mT and is even able to recover some structural details in lower fields. It also assessed sharp features around the lesion area appearing at 37 mT and below in both patients. Fitting took approximately 65 and 150 seconds with the standard method for patient I and II, respectively, whereas the proposed method took 100 and 240 seconds for each patient, respectively.

\begin{figure}[ht]
	\centering
	\includegraphics[width=0.95\textwidth]{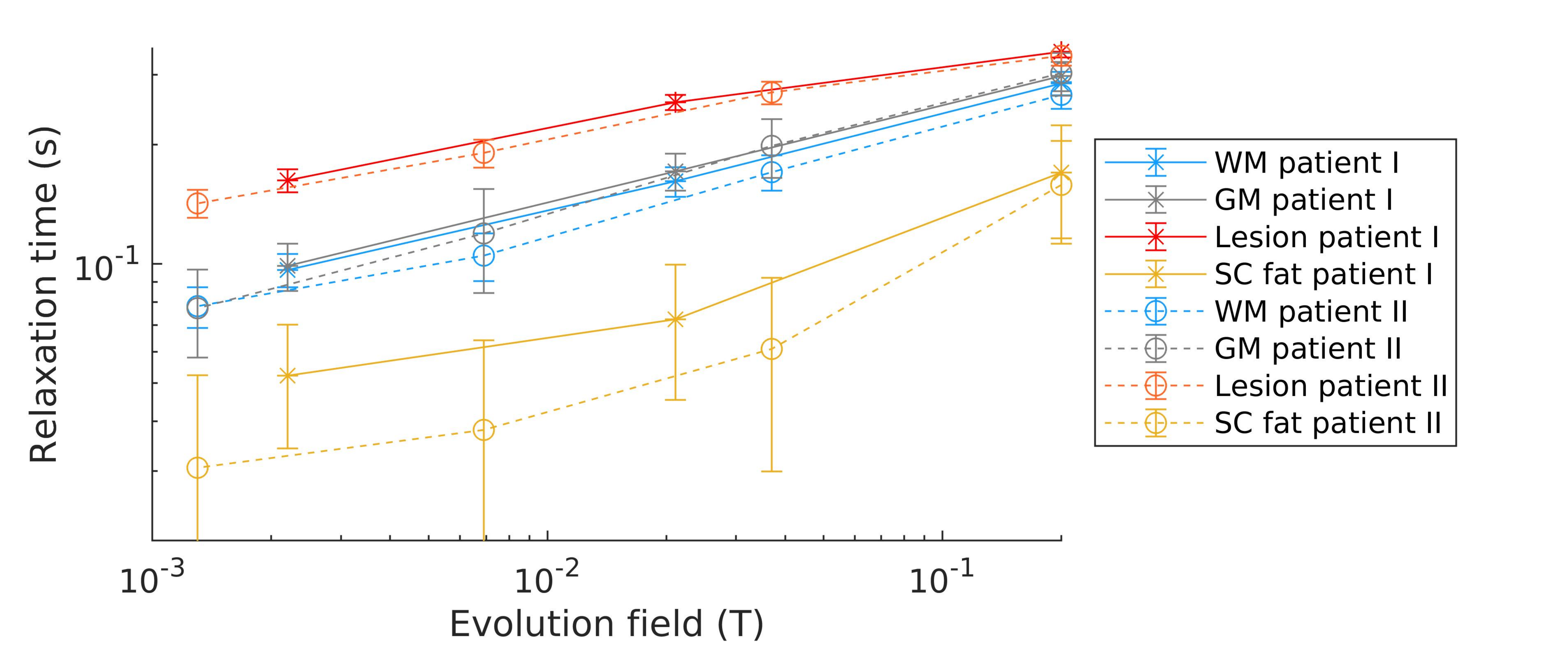}
	\caption{$T_1$ dispersion profiles obtained from the patients I (solid lines) and II (dashed lines) in the regions of subcutaneous fat (SC fat, orange) measured between the scalp and the brain, grey matter (GM, grey) measured over two centimeters of the cortical region, white matter (WM, blue) measured over the inner region of the lobes and the lesion (red). The error bars stand for twice the standard deviation of the $T_1$ values measured across the ROIs. The positioning of the ROIs is depicted in Supporting~Information~Figure~\ref{fig:ROI_patients}.}
	\label{fig:invivo_dispersion_comparison}
\end{figure}

The quality of the $T_1$ maps obtained allowed estimating the $T_1$ dispersion curves for different ROIs, as shown on Figure \ref{fig:invivo_dispersion_comparison} for subcutaneous fat selected under the scalp, the area of the lesion observed at the lowest field strength, and white and grey matter as seen at the highest field strength (the ROIs are shown in Supporting~Information~Figure~\ref{fig:ROI_patients}). The dispersion profiles of fatty tissues show large standard deviations, which may be attributed to the presence of various types of tissues within these ROIs, due to the relatively low resolution of the image. Otherwise, the $T_1$ dispersion profiles of white matter, grey matter, and the areas of the lesions are similar between the two patients. This is encouraging given that the two lesion have a similar diagnosis of ischemic stroke.

\section{Discussion}

The approach used here has high potential to serve as a new standard procedure for fast post-processing of FFC MRI data. As the phantom simulations showed, the noise in the reconstructed $T_1$ maps could be reduced very efficiently while preserving important anatomical details to a high extent. The algorithm outperforms established methods based on pixel-wise fitting of the relaxation profiles yielding lower deviations from the reference values and significantly less variance (Figures~\ref{fig:phantom_T1}~,~\ref{fig:phantom_diff}, and \ref{fig:phantom_T1_2Dhist}). The improved stability results from the combination of information from all acquired fields and exploiting the existence of similar topological structures in the different unknowns. The improved stability is also reflected by increased accuracy of recovered pseudo proton density $C$ and correction factor $\bm{\alpha}$ values (Supporting~Information~Figure~\ref{fig:phantom_alpha_angle}-\ref{fig:phantom_C_abs}). Higher deviations of larger $T_1$ values in the reference methods are due to the employed Tikhonov regularization which penalizes the larger $T_1$ values than lower. Also the $T_1$ maps show significantly reduced variance though there remains some bias which may be, at least partially, due to residual errors in $\bm{\alpha}$. Another remarkable feature of the multi-field methods is their ability to accurately recover the phase information in $C$ and $\bm{\alpha}$, making phase correction prior to fitting obsolete. This in turn can improve $T_1$ maps as no normalization with a noise phase estimate is necessary. 

The advantages of the improved fitting approach become immanent in the in vivo applications (Figures~\ref{fig:invivo_T1_dataset1}~and~\ref{fig:invivo_T1_dataset2}). The standard approaches based on pixel-wise fitting fail to reconstruct image details in both patients. In current practice, k-space windowing filters are applied to recover usable information but this dramatically reduces image resolution by filtering out the high-frequency components of the image, which are responsible for the sharp features. In contrast, the joint regularization approach can recover clearly distinguishable grey and white matter regions at 200 mT on the two patient datasets, previously hidden in noise. The values obtained for the different regions of interest agree well between the patients, given the estimation of the error provided by the variation of the $T_1$ values within each ROI (Figure~\ref{fig:invivo_dispersion_comparison}). The $T_1$ values were systematically higher in patient I than in patient II, which may be attributed to patient variability and different RF receive coil sensitivity relative to the used ROIs. In addition, lesion localization agrees well with conventional MRI and CT based imaging, shown in Supporting~Information~Figure~\ref{fig:conventional}.

As expected from the raw images, the largest $T_1$ contrast for stroke appeared below 0.1 T, where $T_1$ values were larger than that of the surrounding tissues. A cutoff appears between 30 and 100 mT (or equivalently 1.2 to 4.2 MHz) above which the contrast disappears. This is consistent with the fact that higher clinical fields do not show significant $T_1$ changes in acute ischemic stroke. Clearer interpretations may be provided from the analysis of the full data set but a tentative explanations of this phenomenon could be made by taking into account the biological effects of ischemia. During acute ischemia neuron cells swell and burst and this process is likely to disorganise large structures that interact with water over timescales that correspond to the cutoff frequency observed, i.e. between 0.2 and 0.7 ${\mu}s$. The degradation of these components of the brain structure could have the effect to reduce the efficiency of the relaxation pathways at low magnetic fields, as observed here. Another possibility could be that the reduction of water mobility through the membranes of neurons may decrease the contribution of the intracellular water relaxation to the overall signal, which may dominate at low field but could be less efficient at higher fields. These explanation would be consistent with the absence of $T_1$ contrast at higher fields.

As in the phantom images the model-based and spatially regularized methods proved to preserve anatomical features with high spatial frequencies because of using the existence of sharp edges for regularization. These approaches are increasingly accurate with the number of views that can be compared showing the same object, either as a repetition of a recording or as different acquisition of the same field of view, as it is the case here. Hence FFC imaging can benefit from the high information redundancy obtained from the typical acquisition method, which repeats the measurement of the field of view at different evolution times and fields. 

As the proposed approach is model-based, and can therefore provide $T_1$ directly from the raw images, it could be used to reduce the number of steps required to process the image and limits data losses. However model-based approaches also limit the amount of information that is extracted from the image, and properties not covered by the signal equation, may be missed. For instance, brain tissues are known to follow bi-exponential relaxation because of the presence of the myelin sheath around the axons. Hence in a subsequent step, the model will be adapted to the type of scan, or following a test for potential multi-exponential behaviour~\citep{petrov_stapf_2017}. 

Using direct reconstruction from k-space opens up the possibility of undersampled image acquisition while maintaining high quality in the reconstructed $T_1$ maps~\citep{Maier2019}. The proposed method allows for different kinds of undersampling and is not limited to Cartesian sampling or single slice acquisitions. While a single receive coil hast been used for the current study, the extension to a multi-coil setup is straight forward~\citep{Maier2019}. The combination of multiple receive coils and the potential of undersampling k-space could be used to reduce acquisition time in FFC imaging which shall be subject of a future study. The gained acquisition time might lead to a clinical acceptable scan time using the three or four fields shown in this work or could be spent to investigate a multitude of different field strength. However, such extensions would require modifications to the phase correction algorithm which is based on images.

Another advantage of direct reconstruction from k-space data is the validity of the Gaussian noise assumption in the real and complex parts of k-space. In the typically used magnitude images, noise is non-linearly transformed, resulting in a Rician or non-central Chi distribution~\citep{Aja-Fernandez2015}. This invalidates the basic assumptions used to derive the $L^2$-norm data fidelity term and can lead to a bias in the final solution. Even though the data term can be modified to account for these variations the modified version need not be convex or differentiable. Thus, optimization of the correct function might lead to suboptimal solutions or demanding optimization algorithms. In practice, the favourable properties of the $L^2$-norm usually outweigh the drawback of the bias to the theoretically optimal solution and it is thus widely used.

A potential limitation of the proposed approach is the risk of cross-contamination of the information between images due to the joint regularization~\citep{Knoll2017, huber2019, Maier2019}. It is assumed that features share the same edge position. If this assumption is violated in one parameter map, artificial edges might be introduced. The likelihood strongly depends on the used norm for joining the information. As we use a relative weak coupling by means of a Frobenius norm, such cross-contamination is unlikely. It was shown in previous work that Frobenius norm based joint regularization does not show cross-contamination in practice~\citep{Knoll2017, Maier2019, huber2019}. It might only occur if way too strong regularization weights are used, however, such cases would be discarded in practice as images would look unnatural~\citep{Knoll2017}.

The proposed reconstruction and fitting approach is integrated into a recently published Python framework for quantitative MRI~\citep{Maier2020}. This framework allows for an easy adaption to different signal models and thus, a broad application of the proposed method. Adaptions to the signal model can be made by simply editing text files. In addition, 3D regularization strategies are possible which were shown to further improve reconstruction quality~\citep{Maier2019, huber2019}. 

\section{Conclusion}
We have successfully introduced joint TGV\textsuperscript{2} regularization to multi-field $T_1$ quantification from FFC imaging. The highly significant improvements in $T_1$ estimation makes it now possible to obtain clinically usable multi-field $T_1$ maps, and to produce reliable and comparable results. This shows exciting potential for the exploration of low magnetic fields and $T_1$ dispersion effects as illustrated here on two stroke patients.

\subsection{Data Availability Statement}
The Code used for this publication is integrated in PyQMRI~\citep{Maier2020} and is available at \url{https://github.com/IMTtugraz/PyQMRI}. Exemplary data is available at: \url{https://doi.org/10.5281/zenodo.4706998}.

\section*{Acknowledgment}
This article is based upon work from COST Action CA15209, supported by COST (European Cooperation in Science and Technology). Oliver Maier is a Recipient of a DOC Fellowship (24966) of the Austrian Academy  of Sciences at the Institute of Medical Engineering at TU Graz. The authors would like to acknowledge the NVIDIA Corporation Hardware grant support.

\bibliography{references}

\begin{thebibliography}{52}
\providecommand{\natexlab}[1]{#1}
\providecommand{\url}[1]{\texttt{#1}}
\providecommand{\urlprefix}{}

\bibitem[{Steele et~al.(2016)Steele, Rebecca M. and Korb, Jean Pierre and
  Ferrante, Gianni and Bubici, Salvatore}]{Steele2016}
Steele RM, Korb JP, Ferrante G, Bubici S.
\newblock {New applications and perspectives of fast field cycling NMR
  relaxometry}.
\newblock Magnetic Resonance in Chemistry 2016;54(6):502--509.

\bibitem[{Korb(2018)Korb, Jean-Pierre}]{Korb2018}
Korb JP.
\newblock {Multiscale nuclear magnetic relaxation dispersion of complex liquids
  in bulk and confinement}.
\newblock Progress in Nuclear Magnetic Resonance Spectroscopy 2018
  feb;104:12--55.
\newblock
  \urlprefix\url{https://www.sciencedirect.com/science/article/pii/S0079656517300353}.

\bibitem[{Broche et~al.(2011)Broche, Lionel M. and Ismail, Saadiya R. and
  Booth, Nuala A. and Lurie, David J.}]{Broche2011}
Broche LM, Ismail SR, Booth NA, Lurie DJ.
\newblock {Measurement of fibrin concentration by fast field-cycling NMR}.
\newblock Magnetic Resonance in Medicine 2011 oct;67(5):1453--1457.
\newblock \urlprefix\url{https://doi.org/10.1002/mrm.23117}.

\bibitem[{Broche et~al.(2012)Broche, Lionel M. and Ashcroft, George P. and
  Lurie, David J.}]{Broche2012}
Broche LM, Ashcroft GP, Lurie DJ.
\newblock {Detection of osteoarthritis in knee and hip joints by fast
  field-cycling NMR}.
\newblock Magnetic Resonance in Medicine 2012;68(2):358--362.

\bibitem[{Ruggiero et~al.(2018)Ruggiero, Maria Rosaria and Baroni, Simona and
  Pezzana, Stefania and Ferrante, Gianni and {Geninatti Crich}, Simonetta and
  Aime, Silvio}]{Ruggiero2018}
Ruggiero MR, Baroni S, Pezzana S, Ferrante G, {Geninatti Crich} S, Aime S.
\newblock {Evidence for the Role of Intracellular Water Lifetime as a Tumour
  Biomarker Obtained by In Vivo Field-Cycling Relaxometry}.
\newblock Angewandte Chemie International Edition 2018 jun;57(25):7468--7472.
\newblock \urlprefix\url{https://doi.org/10.1002/anie.201713318}.

\bibitem[{{Di Gregorio} et~al.(2019){Di Gregorio}, Enza and Ferrauto, Giuseppe
  and Lanzardo, Stefania and Gianolio, Eliana and Aime,
  Silvio}]{DiGregorio2019}
{Di Gregorio} E, Ferrauto G, Lanzardo S, Gianolio E, Aime S.
\newblock {Use of FCC-NMRD relaxometry for early detection and characterization
  of ex-vivo murine breast cancer}.
\newblock Scientific Reports 2019;9(1):4624.
\newblock \urlprefix\url{https://doi.org/10.1038/s41598-019-41154-9}.

\bibitem[{Lurie et~al.(2010)Lurie, David J. and Aime, Silvio and Baroni, Simona
  and Booth, Nuala A. and Broche, Lionel M. and Choi, Chang Hoon and Davies,
  Gareth R. and Ismail, Saadiya and O'H{\'{o}}g{\'{a}}in, Dara and Pine, Kerrin
  J.}]{Lurie2010}
Lurie DJ, Aime S, Baroni S, Booth NA, Broche LM, Choi CH, et~al.
\newblock {Fast Field-Cycling Magnetic Resonance Imaging}.
\newblock Comptes Rendus Physique 2010;11(2):136--148.

\bibitem[{B{\"{o}}denler et~al.(2019)B{\"{o}}denler, Markus and de Rochefort,
  Ludovic and Ross, P James and Chanet, Nicolas and Guillot, Genevi{\`{e}}ve
  and Davies, Gareth R and G{\"{o}}sweiner, Christian and Scharfetter, Hermann
  and Lurie, David J and Broche, Lionel M}]{Bodenler2019}
B{\"{o}}denler M, de~Rochefort L, Ross PJ, Chanet N, Guillot G, Davies GR,
  et~al.
\newblock {Comparison of fast field-cycling magnetic resonance imaging methods
  and future perspectives}.
\newblock Molecular Physics 2019 apr;117(7-8):832--848.
\newblock \urlprefix\url{https://doi.org/10.1080/00268976.2018.1557349}.

\bibitem[{Hoelscher et~al.(2012)Hoelscher, Uvo Christoph and Lother, Steffen
  and Fidler, Florian and Blaimer, Martin and Jakob, Peter}]{Hoelscher2012}
Hoelscher UC, Lother S, Fidler F, Blaimer M, Jakob P.
\newblock {Quantification and localization of contrast agents using delta
  relaxation enhanced magnetic resonance at 1.5 T}.
\newblock Magnetic Resonance Materials in Physics, Biology and Medicine
  2012;25(3):223--231.

\bibitem[{Harris et~al.(2014)Harris, Chad T. and Handler, William B. and Araya,
  Yonathan and Mart{\'{i}}nez-Santiesteban, Francisco and Alford, Jamu K. and
  Dalrymple, Brian and {Van Sas}, Frank and Chronik, Blaine A. and Scholl,
  Timothy J.}]{Harris2014}
Harris CT, Handler WB, Araya Y, Mart{\'{i}}nez-Santiesteban F, Alford JK,
  Dalrymple B, et~al.
\newblock {Development and optimization of hardware for delta relaxation
  enhanced MRI}.
\newblock Magnetic Resonance in Medicine 2014 oct;72(4):1182--1190.
\newblock \urlprefix\url{http://doi.wiley.com/10.1002/mrm.25014}.

\bibitem[{Chanet et~al.(2020)Chanet, Nicolas and Guillot, Genevi{\`{e}}ve and
  Willoquet, Georges and Jourdain, Laur{\`{e}}ne and Dubuisson, Rose-Marie and
  Reganha, Ga{\"{e}}l and de Rochefort, Ludovic}]{Chanet2020}
Chanet N, Guillot G, Willoquet G, Jourdain L, Dubuisson RM, Reganha G, et~al.
\newblock {Design of a fast field-cycling magnetic resonance imaging system,
  characterization and methods for relaxation dispersion measurements around
  1.5 T}.
\newblock Review of Scientific Instruments 2020 feb;91(2):24102.
\newblock \urlprefix\url{https://doi.org/10.1063/1.5128851}.

\bibitem[{B{\"{o}}denler et~al.(2018)B{\"{o}}denler, Markus and Basini, Martina
  and Casula, Maria Francesca and Umut, Evrim and G{\"{o}}sweiner, Christian
  and Petrovic, Andreas and Kruk, Danuta and Scharfetter,
  Hermann}]{Bodenler2018}
B{\"{o}}denler M, Basini M, Casula MF, Umut E, G{\"{o}}sweiner C, Petrovic A,
  et~al.
\newblock {R1 dispersion contrast at high field with fast field-cycling MRI}.
\newblock Journal of Magnetic Resonance 2018;290:68--75.

\bibitem[{Alford et~al.(2009)Alford, Jamu K. and Rutt, Brian K. and Scholl,
  Timothy J. and Handler, William B. and Chronik, Blaine A.}]{Alford2009}
Alford JK, Rutt BK, Scholl TJ, Handler WB, Chronik BA.
\newblock {Delta relaxation enhanced mr: Improving activation - Speeificity of
  molecular probes through R 1 dispersion imaging}.
\newblock Magnetic Resonance in Medicine 2009;61(4):796--802.

\bibitem[{Araya et~al.(2017)Araya, Yonathan T. and Mart{\'{i}}nez-Santiesteban,
  Francisco and Handler, William B. and Harris, Chad T. and Chronik, Blaine A.
  and Scholl, Timothy J.}]{Araya2017}
Araya YT, Mart{\'{i}}nez-Santiesteban F, Handler WB, Harris CT, Chronik BA,
  Scholl TJ.
\newblock {Nuclear magnetic relaxation dispersion of murine tissue for
  development of T$_1$ ( R$_1$ ) dispersion contrast imaging}.
\newblock NMR in Biomedicine 2017 dec;30(12):e3789.
\newblock \urlprefix\url{http://doi.wiley.com/10.1002/nbm.3789}.

\bibitem[{B{\"{o}}denler et~al.(2019)B{\"{o}}denler, Markus and Malikidogo,
  Kyangwi P and Morfin, Jean-Fran{\c{c}}ois and Aigner, Christoph Stefan and
  T{\'{o}}th, {\'{E}}va and Bonnet, C{\'{e}}lia S and Scharfetter,
  Hermann}]{Bodenler2019a}
B{\"{o}}denler M, Malikidogo KP, Morfin JF, Aigner CS, T{\'{o}}th {\'{E}},
  Bonnet CS, et~al.
\newblock {High-Field Detection of Biomarkers with Fast Field-Cycling MRI: The
  Example of Zinc Sensing}.
\newblock Chemistry - A European Journal 2019 jun;25(35):8236--8239.
\newblock \urlprefix\url{https://doi.org/10.1002/chem.201901157}.

\bibitem[{Carlson et~al.(1992)Carlson, J W and Goldhaber, D M and Brito, A and
  Kaufman, L}]{Carlson1992}
Carlson JW, Goldhaber DM, Brito A, Kaufman L.
\newblock {MR relaxometry imaging. Work in progress.}
\newblock Radiology 1992 sep;184(3):635--639.
\newblock \urlprefix\url{https://doi.org/10.1148/radiology.184.3.1509044}.

\bibitem[{Lurie et~al.(1998)Lurie, David J and Foster, Margaret A and Yeung,
  David and Hutchison, James M S}]{Lurie1998}
Lurie DJ, Foster MA, Yeung D, Hutchison JMS.
\newblock {Design, construction and use of a large-sample field-cycled PEDRI
  imager}.
\newblock Physics in Medicine {\&} Biology 1998;43(7):1877.
\newblock \urlprefix\url{http://stacks.iop.org/0031-9155/43/i=7/a=008}.

\bibitem[{Ungersma et~al.(2006)Ungersma, Sharon E and Matter, Nathaniel I and
  Hardy, Jonathan W and Venook, Ross D and Macovski, Albert and Conolly, Steven
  M and Scott, Greig C}]{Ungersma2006}
Ungersma SE, Matter NI, Hardy JW, Venook RD, Macovski A, Conolly SM, et~al.
\newblock {Magnetic resonance imaging with T1 dispersion contrast.}
\newblock Magnetic Resonance in Medicine 2006;55(6):1362--71.
\newblock \urlprefix\url{http://www.ncbi.nlm.nih.gov/pubmed/16673360}.

\bibitem[{Pine et~al.(2014)Pine, Kerrin J. and Goldie, Fred and Lurie, David
  J.}]{Pine2014}
Pine KJ, Goldie F, Lurie DJ.
\newblock {In vivo field-cycling relaxometry using an insert coil for magnetic
  field offset}.
\newblock Magnetic Resonance in Medicine 2014;72(5):1492--1497.

\bibitem[{Romero et~al.(2020)Romero, Javier A. and Rodriguez, Gonzalo G. and
  Anoardo, Esteban}]{Romero2020}
Romero JA, Rodriguez GG, Anoardo E.
\newblock {A fast field-cycling MRI relaxometer for physical contrasts design
  and pre-clinical studies in small animals}.
\newblock Journal of Magnetic Resonance 2020;311:106682.
\newblock \urlprefix\url{https://doi.org/10.1016/j.jmr.2019.106682}.

\bibitem[{Broche et~al.(2019)Broche, Lionel M and Ross, P James and Davies,
  Gareth R and MacLeod, Mary-Joan and Lurie, David J}]{Broche2019}
Broche LM, Ross PJ, Davies GR, MacLeod MJ, Lurie DJ.
\newblock {A whole-body Fast Field-Cycling scanner for clinical molecular
  imaging studies}.
\newblock Scientific Reports 2019;9(1):10402.
\newblock \urlprefix\url{https://doi.org/10.1038/s41598-019-46648-0}.

\bibitem[{Masiewicz et~al.(2020)Masiewicz, Elzbieta and Ashcroft, George P. and
  Boddie, David and Dundas, Sinclair R. and Kruk, Danuta and Broche, Lionel
  M.}]{masiewicz2020}
Masiewicz E, Ashcroft GP, Boddie D, Dundas SR, Kruk D, Broche LM.
\newblock Towards applying NMR relaxometry as a diagnostic tool for bone and
  soft tissue sarcomas: a pilot study.
\newblock Scientific Reports 2020;10(1).

\bibitem[{Aime et~al.(2018)Aime, Silvio and Botta, Mauro and
  Esteban-G{\'{o}}mez, David and Platas-Iglesias, Carlos}]{Aime2018}
Aime S, Botta M, Esteban-G{\'{o}}mez D, Platas-Iglesias C.
\newblock {Characterisation of magnetic resonance imaging (MRI) contrast agents
  using NMR relaxometry}.
\newblock Molecular Physics 2018;117:898--909.
\newblock
  \urlprefix\url{https://www.tandfonline.com/doi/full/10.1080/00268976.2018.1516898}.

\bibitem[{{Coupe} et~al.(2008)P. {Coupe} and P. {Yger} and S. {Prima} and P.
  {Hellier} and C. {Kervrann} and C. {Barillot}}]{coupe2008}
{Coupe} P, {Yger} P, {Prima} S, {Hellier} P, {Kervrann} C, {Barillot} C.
\newblock An Optimized Blockwise Nonlocal Means Denoising Filter for 3-D
  Magnetic Resonance Images.
\newblock IEEE Transactions on Medical Imaging 2008;27(4):425--441.

\bibitem[{Anand and Sahambi(2010)C. Shyam Anand and Jyotinder S.
  Sahambi}]{ANAND2010842}
Anand CS, Sahambi JS.
\newblock Wavelet domain non-linear filtering for MRI denoising.
\newblock Magnetic Resonance Imaging 2010;28(6):842--861.
\newblock
  \urlprefix\url{https://www.sciencedirect.com/science/article/pii/S0730725X10000767}.

\bibitem[{{Fessler}(2010)J. A. {Fessler}}]{Fessler2010}
{Fessler} JA.
\newblock Model-Based Image Reconstruction for MRI.
\newblock IEEE Signal Processing Magazine 2010;27(4):81--89.

\bibitem[{Knoll et~al.(2011)Knoll, Florian and Bredies, Kristian and Pock,
  Thomas and Stollberger, Rudolf}]{Knoll2011}
Knoll F, Bredies K, Pock T, Stollberger R.
\newblock {Second order total generalized variation (TGV) for MRI}.
\newblock Magnetic Resonance in Medicine 2011;65(2):480--491.
\newblock \urlprefix\url{http:https://doi.org/10.1002/mrm.22595}.

\bibitem[{{Lingala} et~al.(2011)S. G. {Lingala} and Y. {Hu} and E. {DiBella}
  and M. {Jacob}}]{ktslr2011}
{Lingala} SG, {Hu} Y, {DiBella} E, {Jacob} M.
\newblock Accelerated Dynamic MRI Exploiting Sparsity and Low-Rank Structure:
  k-t SLR.
\newblock IEEE Transactions on Medical Imaging 2011;30(5):1042--1054.

\bibitem[{Schloegl et~al.(2017)Schloegl, Matthias and Holler, Martin and
  Schwarzl, Andreas and Bredies, Kristian and Stollberger,
  Rudolf}]{schloegl2017}
Schloegl M, Holler M, Schwarzl A, Bredies K, Stollberger R.
\newblock Infimal convolution of total generalized variation functionals for
  dynamic MRI.
\newblock Magnetic Resonance in Medicine 2017;78(1):142--155.
\newblock
  \urlprefix\url{https://onlinelibrary.wiley.com/doi/abs/10.1002/mrm.26352}.

\bibitem[{Doneva et~al.(2010)Doneva, Mariya and Börnert, Peter and Eggers,
  Holger and Stehning, Christian and Sénégas, Julien and Mertins,
  Alfred}]{doneva2010}
Doneva M, Börnert P, Eggers H, Stehning C, Sénégas J, Mertins A.
\newblock Compressed sensing reconstruction for magnetic resonance parameter
  mapping.
\newblock Magnetic Resonance in Medicine 2010;64(4):1114--1120.
\newblock
  \urlprefix\url{https://onlinelibrary.wiley.com/doi/abs/10.1002/mrm.22483}.

\bibitem[{Sumpf et~al.(2011)Sumpf, Tilman J. and Uecker, Martin and Boretius,
  Susann and Frahm, Jens}]{sumpf2011}
Sumpf TJ, Uecker M, Boretius S, Frahm J.
\newblock Model-based nonlinear inverse reconstruction for T2 mapping using
  highly undersampled spin-echo MRI.
\newblock Journal of Magnetic Resonance Imaging 2011;34(2):420--428.
\newblock
  \urlprefix\url{https://onlinelibrary.wiley.com/doi/abs/10.1002/jmri.22634}.

\bibitem[{Wang et~al.(2018)Wang, Xiaoqing and Roeloffs, Volkert and Klosowski,
  Jakob and Tan, Zhengguo and Voit, Dirk and Uecker, Martin and Frahm,
  Jens}]{Wang2017}
Wang X, Roeloffs V, Klosowski J, Tan Z, Voit D, Uecker M, et~al.
\newblock Model-based T1 mapping with sparsity constraints using single-shot
  inversion-recovery radial FLASH.
\newblock Magnetic Resonance in Medicine 2018;79(2):730--740.
\newblock
  \urlprefix\url{https://onlinelibrary.wiley.com/doi/abs/10.1002/mrm.26726}.

\bibitem[{Maier et~al.(2019)Maier, Oliver and Schoormans, Jasper and Schloegl,
  Matthias and Strijkers, Gustav J and Lesch, Andreas and Benkert, Thomas and
  Block, Tobias and Coolen, Bram F and Bredies, Kristian and Stollberger,
  Rudolf}]{Maier2019}
Maier O, Schoormans J, Schloegl M, Strijkers GJ, Lesch A, Benkert T, et~al.
\newblock {Rapid T1 quantification from high resolution 3D data with
  model-based reconstruction}.
\newblock Magnetic Resonance in Medicine 2019 mar;81(3):2072--2089.
\newblock \urlprefix\url{https://doi.org/10.1002/mrm.27502}.

\bibitem[{Rudin et~al.(1992)Rudin, Leonid I and Osher, Stanley and Fatemi,
  Emad}]{Rudin1992}
Rudin LI, Osher S, Fatemi E.
\newblock {Nonlinear total variation based noise removal algorithms}.
\newblock Physica D: Nonlinear Phenomena 1992;60(1):259--268.
\newblock
  \urlprefix\url{http://www.sciencedirect.com/science/article/pii/016727899290242F}.

\bibitem[{Bredies et~al.(2010)Bredies, Kristian and Kunisch, Karl and Pock,
  Thomas}]{Bredies2010}
Bredies K, Kunisch K, Pock T.
\newblock {Total Generalized Variation}.
\newblock SIAM Journal on Imaging Sciences 2010 jan;3(3):492--526.
\newblock \urlprefix\url{https://doi.org/10.1137/090769521}.

\bibitem[{Lustig et~al.(2007)Lustig, Michael and Donoho, David and Pauly, John
  M}]{Lustig2007}
Lustig M, Donoho D, Pauly JM.
\newblock {Sparse MRI: The application of compressed sensing for rapid MR
  imaging}.
\newblock Magnetic Resonance in Medicine 2007 dec;58(6):1182--1195.
\newblock \urlprefix\url{https://doi.org/10.1002/mrm.21391}.

\bibitem[{Wang et~al.(2018)Wang, Xiaoqing and Roeloffs, Volkert and Klosowski,
  Jakob and Tan, Zhengguo and Voit, Dirk and Uecker, Martin and Frahm,
  Jens}]{Wang2018}
Wang X, Roeloffs V, Klosowski J, Tan Z, Voit D, Uecker M, et~al.
\newblock {Model-based T1 mapping with sparsity constraints using single-shot
  inversion-recovery radial FLASH}.
\newblock Magnetic Resonance in Medicine 2018 feb;79(2):730--740.
\newblock \urlprefix\url{https://doi.org/10.1002/mrm.26726}.

\bibitem[{Lai et~al.(2018)Lai, Zongying and Zhang, Xinlin and Guo, Di and Du,
  Xiaofeng and Yang, Yonggui and Guo, Gang and Chen, Zhong and Qu,
  Xiaobo}]{Lai2018}
Lai Z, Zhang X, Guo D, Du X, Yang Y, Guo G, et~al.
\newblock {Joint sparse reconstruction of multi-contrast MRI images with graph
  based redundant wavelet transform}.
\newblock BMC Medical Imaging 2018;18(1):7.
\newblock \urlprefix\url{https://doi.org/10.1186/s12880-018-0251-y}.

\bibitem[{Bredies and Holler(2020)Kristian Bredies and Martin
  Holler}]{Bredies2020}
Bredies K, Holler M.
\newblock Higher-order total variation approaches and generalisations.
\newblock Inverse Problems 2020 Dec;36(12):123001.
\newblock \urlprefix\url{https://doi.org/10.1088/1361-6420/ab8f80}.

\bibitem[{H{\'{o}}g{\'{a}}in et~al.(2010)Dara {\'{O}} H{\'{o}}g{\'{a}}in and
  Gareth R Davies and Simona Baroni and Silvio Aime and David J
  Lurie}]{signaleq}
H{\'{o}}g{\'{a}}in D{\'{O}}, Davies GR, Baroni S, Aime S, Lurie DJ.
\newblock The use of contrast agents with fast field-cycling magnetic resonance
  imaging.
\newblock Physics in Medicine and Biology 2010 nov;56(1):105--115.
\newblock \urlprefix\url{https://doi.org/10.1088/0031-9155/56/1/007}.

\bibitem[{Bredies(2014)Bredies, Kristian}]{Bredies2014}
Bredies K.
\newblock {Recovering Piecewise Smooth Multichannel Images by Minimization of
  Convex Functionals with Total Generalized Variation Penalty BT - Efficient
  Algorithms for Global Optimization Methods in Computer Vision}.
\newblock Berlin, Heidelberg: Springer Berlin Heidelberg; 2014. p. 44--77.

\bibitem[{Knoll et~al.(2017)Knoll, F and Holler, M and Koesters, T and Otazo, R
  and Bredies, K and Sodickson, D K}]{Knoll2017}
Knoll F, Holler M, Koesters T, Otazo R, Bredies K, Sodickson DK.
\newblock {Joint MR-PET Reconstruction Using a Multi-Channel Image
  Regularizer}.
\newblock IEEE Transactions on Medical Imaging 2017;36(1):1--16.

\bibitem[{Kaltenbacher and Hofmann(2010)Barbara Kaltenbacher and Bernd
  Hofmann}]{Kaltenbacher2010}
Kaltenbacher B, Hofmann B.
\newblock Convergence rates for the iteratively regularized
  Gauss{\textendash}Newton method in Banach spaces.
\newblock Inverse Problems 2010 feb;26(3):035007.
\newblock \urlprefix\url{https://doi.org/10.1088/0266-5611/26/3/035007}.

\bibitem[{Chambolle and Pock(2011)Chambolle, Antonin and Pock,
  Thomas}]{Chambolle2011}
Chambolle A, Pock T.
\newblock {A First-Order Primal-Dual Algorithm for Convex Problems with
  Applications to Imaging}.
\newblock Journal of Mathematical Imaging and Vision 2011;40(1):120--145.
\newblock \urlprefix\url{https://doi.org/10.1007/s10851-010-0251-1}.

\bibitem[{Malitsky and Pock(2018)Malitsky, Yura and Pock,
  Thomas}]{Malitsky2018}
Malitsky Y, Pock T.
\newblock {A First-Order Primal-Dual Algorithm with Linesearch}.
\newblock SIAM Journal on Optimization 2018 jan;28(1):411--432.
\newblock \urlprefix\url{https://doi.org/10.1137/16M1092015}.

\bibitem[{Ross et~al.(2015)Ross, P. James and Broche, Lionel M. and Lurie,
  David J.}]{Ross2014}
Ross PJ, Broche LM, Lurie DJ.
\newblock Rapid field-cycling MRI using fast spin-echo.
\newblock Magnetic Resonance in Medicine 2015;73(3):1120--1124.
\newblock
  \urlprefix\url{https://onlinelibrary.wiley.com/doi/abs/10.1002/mrm.25233}.

\bibitem[{Broche et~al.(2017)Broche, Lionel M. and Ross, P. James and Davies,
  Gareth R. and Lurie, David J.}]{broche2017}
Broche LM, Ross PJ, Davies GR, Lurie DJ.
\newblock Simple algorithm for the correction of MRI image artefacts due to
  random phase fluctuations.
\newblock Magnetic Resonance Imaging 2017;44:55–59.

\bibitem[{Maier et~al.(2020)Oliver Maier and Stefan M. Spann and Markus
  Bödenler and Rudolf Stollberger}]{Maier2020}
Maier O, Spann SM, Bödenler M, Stollberger R.
\newblock PyQMRI: An accelerated Python based Quantitative MRI toolbox.
\newblock Journal of Open Source Software 2020;5(56):2727.
\newblock \urlprefix\url{https://doi.org/10.21105/joss.02727}.

\bibitem[{Jin and Zhong(2013)Jin, Qinian and Zhong, Min}]{jin_zhong_2013}
Jin Q, Zhong M.
\newblock On the iteratively regularized Gauss–Newton method in Banach spaces
  with applications to parameter identification problems.
\newblock Numerische Mathematik 2013;124(4):647–683.

\bibitem[{Petrov and Stapf(2017)Petrov, Oleg V. and Stapf,
  Siegfried}]{petrov_stapf_2017}
Petrov OV, Stapf S.
\newblock Parameterization of NMR relaxation curves in terms of logarithmic
  moments of the relaxation time distribution.
\newblock Journal of Magnetic Resonance 2017;279:29–38.

\bibitem[{Aja-Fern{\'{a}}ndez and Trist{\'{a}}n-Vega(2015)Aja-Fern{\'{a}}ndez,
  Santiago and Trist{\'{a}}n-Vega, Antonio}]{Aja-Fernandez2015}
Aja-Fern{\'{a}}ndez S, Trist{\'{a}}n-Vega A.
\newblock {A review on statistical noise models for Magnetic Resonance
  Imaging}.
\newblock LPI, ETSI Telecomunicacion, Univ Valladolid, Spain, Tech Rep 2015;.

\bibitem[{Huber et~al.(2019)Richard Huber and Georg Haberfehlner and Martin
  Holler and Gerald Kothleitner and Kristian Bredies}]{huber2019}
Huber R, Haberfehlner G, Holler M, Kothleitner G, Bredies K.
\newblock Total generalized variation regularization for multi-modal electron
  tomography.
\newblock Nanoscale 2019;11(12):5617--5632.
\newblock \urlprefix\url{https://doi.org/10.1039/c8nr09058k}.

\end{thebibliography}

\clearpage
\listoffigures
\listoftables
\listofsuppfigures
\clearpage

\begin{table}[ht]
\centering
\caption{Evolution times and fields used to generate the simulated images and corresponding timings for the in vivo acquisitions.}
\label{tab:table-simlations}
\begin{tabular}{cccccccc}
Application &Variable      & Field (mT) & \multicolumn{5}{c}{Evolution time (ms)}  \\ \hline
\cellcolor{white}&$t_{n}^{evo_1}$ & 200        & 455 & 242 & 129 &  68 &  36 \\ 
\cellcolor{white}&$t_{n}^{evo_2}$ & 21.1       & 282 & 150 &  80 &  42 &  23 \\ 
\multirow{-3}{*}{\shortstack[1]{Simulation}}&$t_{n}^{evo_3}$ & 2.2        & 136 &  73 &  39 &  21 &  11 \\ 
\cellcolor{black!10}&$t_{n}^{evo_1}$ & 200        & 455 & 242 & 129 &  68 &  36 \\ 
\cellcolor{black!10}&$t_{n}^{evo_2}$ & 21.1       & 282 & 150 &  80 &  42 &  23 \\ 
\multirow{-3}{*}{\cellcolor{black!10}\shortstack[1]{Patient I}}&$t_{n}^{evo_3}$ & 2.2        & 136 &  73 &  39 &  21 &  11 \\ 
\cellcolor{white}&$t_{n}^{evo_1}$ & 200        & 455 & 196 &  84 &  36 & \\ 
\cellcolor{white}&$t_{n}^{evo_2}$ & 37       & 338 & 145 &  63 &  27 &   \\ 
\cellcolor{white}&$t_{n}^{evo_3}$ & 6.9       & 196 & 84 &  36 &  16 &  \\ 
\multirow{-4}{*}{\cellcolor{white}\shortstack[1]{Patient II}}&$t_{n}^{evo_4}$ & 1.3 & 114 &  49 &  21 &  9 &  \\ 
\end{tabular}
\end{table}

\begin{table}[ht]
\centering
\caption{Parameter values selected to generate the simulated images.}
\label{tab:table-T1-simulations}
\begin{tabular}{cccccc}
 & ROI & 1 & 2 & 3 & 4 \\
Parameter & & & & &  \\\hline
\cellcolor{white} & a & 5.6 & 4.4 & 2.6 & 3.8 \\
\multirow{-2}{*}{power law} & b & -0.1 & -0.15 & -0.3 & -0.08 \\
\cellcolor{black!10} &at 200 mT & 152 & 178.5 & 237.3 & 231.4\\
\cellcolor{black!10} &at 21 mT & 121.3 & 127.3 & 120.7 & 193.2\\
\multirow{-3}{*}{$T_1$ (ms)}& at 2.2 mT & 96.8 & 90.8 & 61.3 & 161.3\\
\cellcolor{white} &at 200 mT & \multicolumn{4}{c}{1/0.5236}\\
\cellcolor{white} &at 21 mT & \multicolumn{4}{c}{0.75/0.6981}\\
\multirow{-3}{*}{\shortstack[1]{$\alpha$ \\abs (a.u.)/phase (rad)}}& at 2.2 mT & \multicolumn{4}{c}{0.6/0.8727}\\
{$C$ (a.u.)}& at all field & 1 & 1/3 & 2/3 & 2.03/3\\
\end{tabular}
\end{table}



\clearpage
\clearpage
\section*{Supporting Information}
\setcounter{page}{1}
\pagenumbering{Roman}
\captionsetup[suppfigure]{name=Supporting Information Figure}
\appendix
\section{Mathematical derivations}
\label{sec:appA}
The main computational burden lies in the optimization of the convex inner problems of each GN iteration. Each consists of an iterative solution to the following optimization task
\begin{linenomath*}
\begin{align} \label{eq:linapp}
   \underset{u,v}{\min}\quad 
\frac{1}{2}&\|
\mathbf{DS}|_{u=u^{k}}u-\tilde{\mathbf{d}}^k\|_2^2 
+ \nonumber\\
\gamma_k(
\beta_0&\|\nabla u - v\|_{1,2,F} + \beta_1|\|\mathcal{E}v\|_{1,2,F}) +
\nonumber\\ \frac{\delta_k}{2}&\|u-u^k\|_{M_k}^2.
\end{align}
\end{linenomath*}

The $\|\cdot\|_{1,2,F}$ terms resemble the Frobenius type TGV$^2$ functionals, 
joining common spatial information of the unknown parameter maps and are 
defined as 
\begin{linenomath*}
\begin{equation}
\|v\|_{1,2,F} = \sum _{x,y} \sqrt{ \sum _{l=1}^{N_u} 
|v_{x,y}^{1,l}|^ 2 + 
|v_{x,y}^{2,l}|^ 2}
\end{equation}
\end{linenomath*}
with $v = (v^{1,l},v^{2,l})^{N_u}_{l=1}\in U^{2\times N_u}$ 
constituting the approximation of 2D spatial derivatives using $\nabla$, 
and for the symmetrized gradient $\mathcal{E}$ $\chi = 
(\chi^{1,l},\chi^{2,l},\chi^{3,l})^{N_u}_{l=1} 
\in 
U^{3\times N_u}$
\begin{linenomath*}
\begin{equation}
\|\chi\|_{1,2,F} = \sum _{x,y} \sqrt{ \sum _{l=1}^{N_u} 
|\chi_{x,y}^{1,l}|^ 2 + |\chi_{x,y}^{2,l}|^ 2 + 2|\chi_{x,y}^{3,l}|^ 2}.
\end{equation}
\end{linenomath*}

$\nabla: U^{N_u} \to U^{2\times N_u} $ and $\mathcal{E}: U^{2\times 
N_u} \to U^{3\times N_u}$ are defined as
\begin{linenomath*}
\[
\nabla u = \Big( \delta_{x+} u^l,\delta_{y+} u^l\Big)_{l=1}^{N_u}
\]
\end{linenomath*}
and
\begin{linenomath*}
\begin{align*}
\mathcal{E}v = \Big( 
\delta_{x-}v^{1,l},\delta_{y-}v^{2,l},\frac{\delta_{y-}v^{1,l} + \delta _{x-}v^{2,l}}{2}
\Big)_{l=1}^{N_u}.
\end{align*}
\end{linenomath*}
The gradient $\nabla$ and symmetrized gradient $\mathcal{E}$ operations are computed by taking the finite differences, defined as forward $\delta_{x+},\,\delta_{y+}$ and backward
$\delta_{x-},\,\delta_{y-}$ differences with respect to the spatial directions along $(x,y)$. At the boundaries, the image is symmetrically extended. To achieve the saddle point formulation
\begin{linenomath*}
\begin{equation}\label{eq:PD_saddle_ap}
\underset{x}{\min}\,\underset{y}{\max}~ \left<\mathrm{K}x,y\right> + G(x) - 
F^*(y),
\end{equation}
\end{linenomath*}
as required for the applied primal-dual algorithm, 
the linearized problem of Eq. \ref{eq:linearized} can be reformulated by applying the convex conjugate:
\begin{linenomath*}
\begin{alignat*}{3}
&  &&\min_{x=(u,v)} ~
&&\frac{1}{2}\|
\mathbf{DS}|_{u=u^{k}}u-\tilde{\mathbf{d}}^k\|_2^2  + \\
& && &&\gamma_k(
\beta_0\|\nabla u - v\|_{1,2,F} + \beta_1|\|\mathcal{E}v\|_{1,2,F}) + \\ 
& && &&\frac{\delta_k}{2}\|u-u^k\|_{M_k}^2\\
&{\Leftrightarrow}
 &&\min_{x} \max_{y=(z_0,z_1,r)}
 &&\left\{\left<\mathbf{DS}|_{u=u^{k}}|_{u=u^{k}} u,\mathbf{r}\right> -
\left<\tilde{\mathbf{d}}^k,\mathbf{r}\right> - \frac{1}{2} \|\mathbf{r} \|_2^2 \right\}+\\
& && &&\left<K_1x,z\right> 
-\mathcal{I}_{\{\|\cdot \|_{\infty \leq \beta_0\gamma_k}\}}(z_0)
-\mathcal{I}_{\{\|\cdot \|_{\infty \leq \beta_1\gamma_k}\}}(z_1) \\
& && && +\frac{\delta_k}{2}\|u-u^{k}\|_{M_k}^2\\
 &{\Leftrightarrow}
  &&\min_{x} \max_{y} ~ &&\left<\mathrm{K}x,y\right> + G(x) - F^*(y). 
\end{alignat*}
\end{linenomath*}
with
\begin{linenomath*}
\begin{align*}
 K = \left( \begin{matrix}
	\mathbf{DS} & 0	\\
 \nabla & -id	\\
 0 & \mathcal{E}	\\
\end{matrix}\right), \quad  K_1 = \left( \begin{matrix}
 \nabla & -id	\\
 0 & \mathcal{E}	\\
\end{matrix}\right), \quad z = (z_0, z_1)^T.
\end{align*} 
\end{linenomath*}
\begin{linenomath*}
\begin{align*}
F^*(y) &= 
\left\{\left<\mathbf{DS}|_{u=u^{k}} u,\mathbf{r}\right> -
\left<\tilde{\mathbf{d}}^k,\mathbf{r}\right>+ \frac{1}{2} \| \mathbf{r} 
\|_2^2\right\} +\mathcal{I}_{\{\|\cdot \|_{\infty \leq \beta_0\gamma_k}\}}(z_0)
+\mathcal{I}_{\{\|\cdot \|_{\infty \leq \beta_1\gamma_k}\}}(z_1), \\
G(x) &=\frac{\delta_k}{2}\| u - u^{k} \|_{M_k}^2. \\
\end{align*} 
\end{linenomath*}
The projection $\mathcal{I}_{\{\|\cdot \|_{\infty \leq \alpha_p\gamma_k}\}}(z_p)$ stems from to the convex conjugate of the $L^1$-norm which amounts to the 
indicator function of the $L^\infty$-norm unit ball, scaled by the corresponding regularization parameter $\alpha_p\gamma_k$
\begin{linenomath*}
\begin{align*}
\mathcal{I}_{\{\|\cdot \|_{\infty \leq \alpha_p\gamma_k}\}}(z_p) = \begin{cases}
0 & \|z_p\|_\infty \leq \alpha_p\gamma_k \\
\infty & else
\end{cases}
\end{align*}
\end{linenomath*}
$\mathbf{DS}|_{u=u^{k}}$ is the Jacobian matrix of $S$ evaluated at 
$u=u^{k}$ of the non-linear FFC signal equation for all fields $\mathbf{B_0}$ and their corresponding inversions times $\mathbf{t}^{E_i}$:
\begin{center}
\begin{linenomath*}
\begin{equation}
  \mathbf{DS}:  u = (u_l)_{l=1}^{N_u}
                \mapsto  \left( \mathcal{F}\sum\limits_{l=1}^{N_u}  \left[ \left. 
\frac{
                	\partial S_{B^{E_i}_0,t^{E_i}_n}}{\partial u_l} \right|_{u = u^{k}}  
                u_l \right] \right)_{n=1}^{N_d}       = (\xi_n)_{n=1}^{N_d}.
\end{equation}
\end{linenomath*}
\end{center}
To update steps of the PD algorithm are defined as
\begin{center}
\begin{linenomath*}
\begin{equation}
\begin{aligned} 
  y^{n+1} &= (id+\sigma \partial F^*)^{-1}(y^n+\sigma K \overline{x}^n)\\
  x^{n+1} &= (id+\tau \partial G)^{-1}(x^n-\tau K^H y^{n+1})\\
  \overline{x}^{n+1} &= x^{n+1} + \theta(x^{n+1} - x^n),
\end{aligned}
\end{equation}
\end{linenomath*}
\end{center}
with $id$ amounting to the identity matrix and $\theta\in[0,1]$.
To compute the updates, additional operations need to be defined, 
which will be covered in the next few paragraphs.

First, the adjoint operations to the linear operator $K$ of the forward problem, termed $K^H$ 
with $^H$ being the Hermitian transpose operation, is defined as
\begin{center}
\begin{linenomath*}
\begin{equation}
\begin{aligned}
  K^H = &\begin{pmatrix}    
            \mathbf{DS}^H & - \text{div}^1 & 0\\
            0 & -id &-\text{div}^2 
            \end{pmatrix},    
\end{aligned}
\end{equation}
\end{linenomath*}
\end{center}
where the divergence operators $\text{div}^1$ and $\text{div}^2$ are the 
negative adjoints of $\nabla $ and $\mathcal{E}$, respectively. 
The adjoint of the Jacobi matrix $\mathbf{DS}^H$ amounts to a simple complex transpose operation
in matrix notation, which can be written in operator notation as
\begin{center}
\begin{linenomath*}
\begin{equation*}
  \mathbf{DS}^H: \xi = (\xi_n)_{n=1}^{N_d} \mapsto 
            \left(  \sum\limits_{n=1}^{N_d}\left. \overline{
              \frac{\partial S_{B^{E_i}_0,t^{E_i}_n}(u)}
              {\partial u_l}} \right|_{u = u^{k}} \mathcal{F}^{H}\xi_n 
                    \right)_{l=1}^{N_u} = (u_l)_{l=1}^{N_u} = u.
\end{equation*}
\end{linenomath*}
\end{center}
Finally, the proximal maps, projecting on the function $F^*$
amount to simple point wise operations
\begin{linenomath*}
\begin{align*}
&P_{\beta_0}(\xi)  = \frac{\xi}{\max \left( 1,  \frac{ 
\vert \xi \vert }{\beta_0\gamma} \right)}, \\ 
&P_{\beta_1}(\xi)  = \frac{\xi}{\max \left( 1,  \frac{ 
\vert \xi \vert }{\beta_1\gamma} \right)}, \\ 
&P_{\sigma L^2}(\xi) = \frac{\xi - \sigma \tilde{d}^k}{1+{\sigma}}
\end{align*}
\end{linenomath*}
and the proximal map of $G$ is defined as
\begin{linenomath*}
\begin{align*}
&P_G(\xi)={(id+{\tau\delta_k M_k})^{-1}}{(\tau\delta_k M_k u^{k}+\xi)}.
\end{align*}
\end{linenomath*}
Making use of the fact that $M_k$ is a diagonal matrix, the inversion of $(id+{\tau\delta_k M_k})$ is trivial and thus $P_G(\xi)$ can be computed in a point-wise fashion.

\clearpage
\section{Pseudo Code}
\begin{algorithm2e}
    \DontPrintSemicolon

    \textbf{Initialize:}  $(u^0,v^0)$, 
$(\overline{u^0},\overline{v^0}),$ 
$(z_0^0,z_1^0,r^0)$, $\tau^0 > 0 $, $\kappa^0 = 1, \theta^0 = 1$, $\mu = 0.5$\\

    \textbf{Iterate:} \\
    \vspace*{1em}
    \textbf{Primal Update: }\\
    \vspace*{0.2em}
    \hspace*{1em} $u^{m+1} \leftarrow P_{\tau^mG}\left( u^m - \tau^m \left( 
-\text{div}^1 z_0^{m}
+ \mathbf{DS}^Hr^{m}  \right)  \right)$ \\
    \hspace*{1em}$v^{m+1} \leftarrow v - \tau^m \left(- \text{div}^2 z_1^{m} - 
z_0^{m}
\right) $ \\
    \textbf{Update $\kappa$ and $\tau$:} \\
    \hspace*{1em}$\kappa^{m+1} \leftarrow \kappa^m(1+\delta_k\tau^m)$\\
    \hspace*{1em}$\tau^{m+1} \leftarrow 
\tau^m\sqrt{\frac{\kappa^m}{\kappa^{m+1}}(1+\theta^m)}$\\
    \vspace*{1em}
    \textbf{Start Linesearch:} \\
   \textbf{	Update $\theta$:} \\
    \hspace*{1em}$\theta^{m+1} \leftarrow \frac{\tau^{m+1}}{\tau^m}$\\
    \vspace*{0.2em}    
    \textbf{Extrapolation:} \\
    \vspace*{0.2em}
    \hspace*{1em}$(\overline{u}^{m+1},\overline{v}^{m+1}) 
    \leftarrow (u^{m+1},v^{m+1}) + \theta^{m+1}
((u^{m+1},v^{m+1})-(u^m,v^m)) $    \\
    \vspace*{0.2em}    
    \textbf{Dual Update:} \\
    \vspace*{0.2em}
    \hspace*{1em}$z_0^{m+1}\leftarrow P_{\beta_0} \left( z_0^m + 
\kappa^{m+1}\tau^{m+1} (\nabla
\overline{u}^{m+1} - \overline{v}^{m+1}) \right)$ \\
    \hspace*{1em}$z_1^{m+1} \leftarrow P_{\beta_1} \left( z_1^m + 
\kappa^{m+1}\tau^{m+1} 
(\mathcal{E}\overline{v}^{m+1}) \right)$ \\
    \hspace*{1em}$r^{m+1} \leftarrow P_{\kappa^{m+1}\tau^{m+1}L^2} \left( r^m + 
\kappa\tau^{m+1}(\mathbf{DS}\,\overline{u}^{m+1})\right)$ \\
    \vspace*{0.2em}
    \textbf{break Linesearch if:} \\
    \vspace*{0.2em}
    \hspace*{1em}$\sqrt{\kappa^{m+1}}\tau^{m+1}\|K^H\,y^{m+1}-K^H\,y^m\| 
\leq \|y^{m+1}-y^m\|$\\
    \vspace*{0.2em}
    \textbf{else:} \\
    \vspace*{0.2em}    
    \hspace*{1em}$\tau^{m+1} \leftarrow \tau^{m+1}\mu $\\
    \vspace*{1em}
    \textbf{Update:}\\
    \vspace*{0.2em}        
    \hspace*{1em}$(u^m,v^m,\tau^m) \leftarrow 
(u^{m+1},v^{m+1},\tau^{m+1}) $\\
    \hspace*{1em}$(z_0^m,z_1^m,r^m) \leftarrow (z_0^{m+1},z_1^{m+1},r^{m+1})$
    \vspace*{1em}    
    \caption{Pseudo code for the solution of the inner problems of each GN step based on the primal-dual algorithm with linesearch. As described in the original publication, linearity of certain operations can be used to speed up the computation of the linesearch procedure.} 
    \label{alg:tgvsolve}
\end{algorithm2e} 

\begin{suppfigure}[ht]
	\centering
	\includegraphics[width=0.95\textwidth]{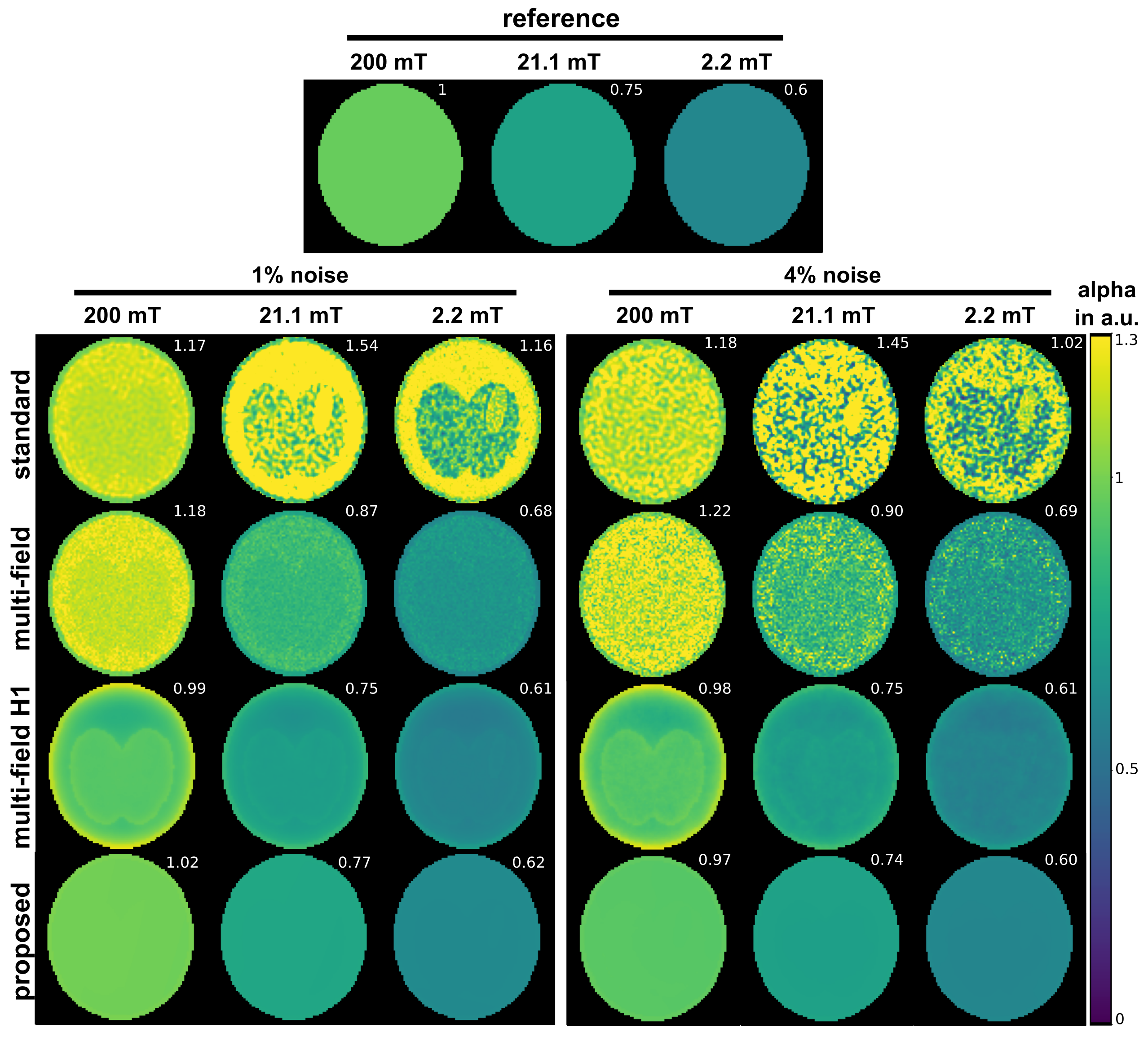}
	\caption{Absolute value of multi-field $\alpha$ maps obtained from simulated FFC imaging inversion recovery data. The reference $\alpha$ maps for three different evolution fields (200~mT, 21.1~mT and 2.2~mT) are shown at the top.
	The different reconstruction methods are presented in each row. Standard refers to single field pixel-wise fitting, multi-field to combined field, pixel-wise fitting approach and H1 to the model-based approach with regularization using the squared $L^2$-norm of the gradient. The proposed method is shown in the last row. The columns show increasing noise from left to right. Values next to each figure represent the mean value within the simulated phantom in a.u..}
	\label{fig:phantom_alpha_abs}
\end{suppfigure}
\begin{suppfigure}[ht]
	\centering
	\includegraphics[width=0.95\textwidth]{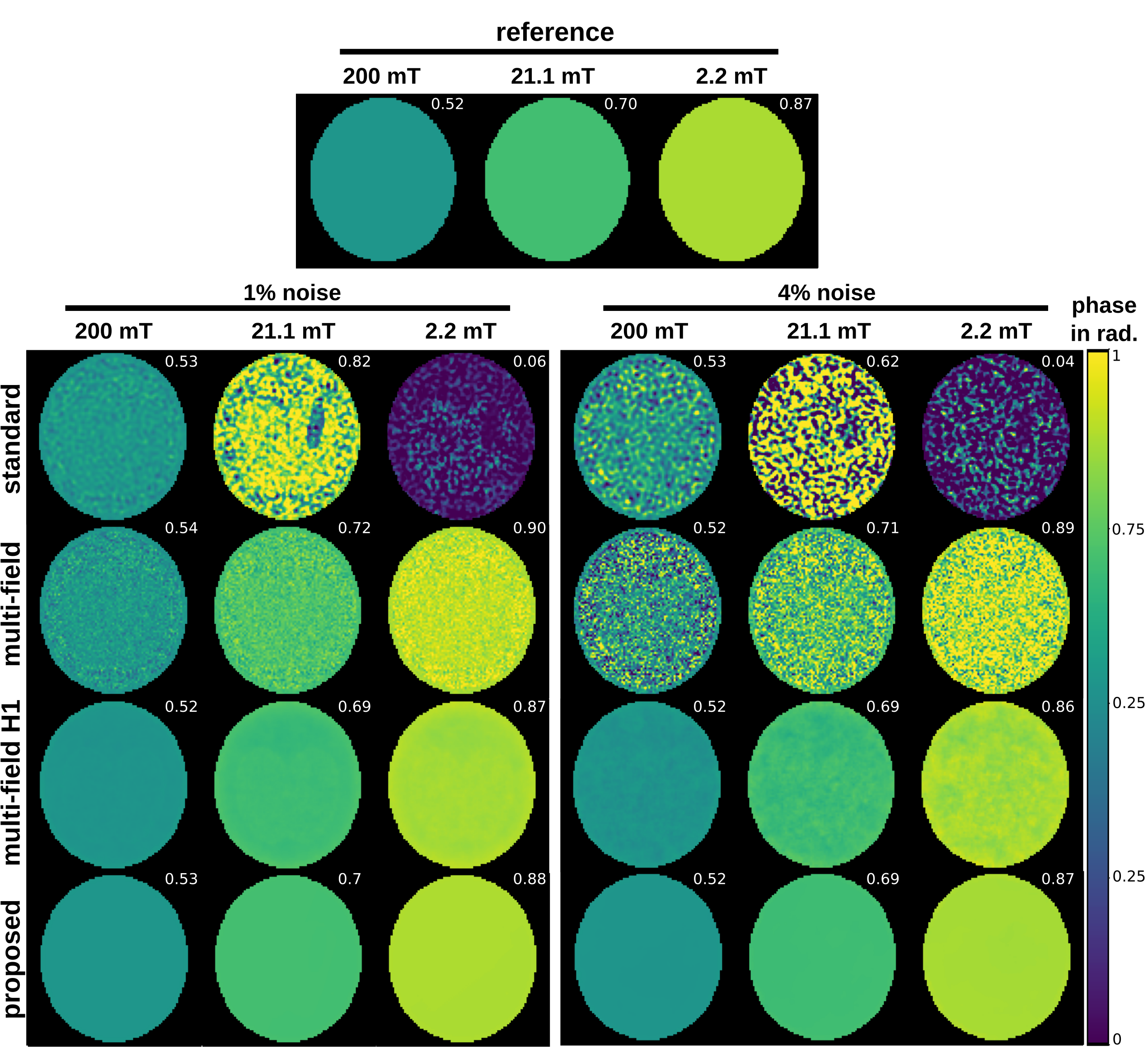} 
	\caption{Phase of multi-field $\alpha$ maps obtained from simulated FFC imaging inversion recovery data. The reference phase maps for three different evolution fields (200~mT, 21.1~mT and 2.2~mT) are shown at the top.
	The different reconstruction methods are presented in each row. Standard refers to single field pixel-wise fitting, multi-field to combined field, pixel-wise fitting approach and H1 to the model-based approach with regularization using the squared $L^2$-norm of the gradient. The proposed method is shown in the last row. The columns show increasing noise from left to right.
	Values next to each figure represent the mean value within the simulated phantom in radiant.}
	\label{fig:phantom_alpha_angle}
\end{suppfigure}
\begin{suppfigure}[ht]
	\centering
	\includegraphics[width=0.95\textwidth]{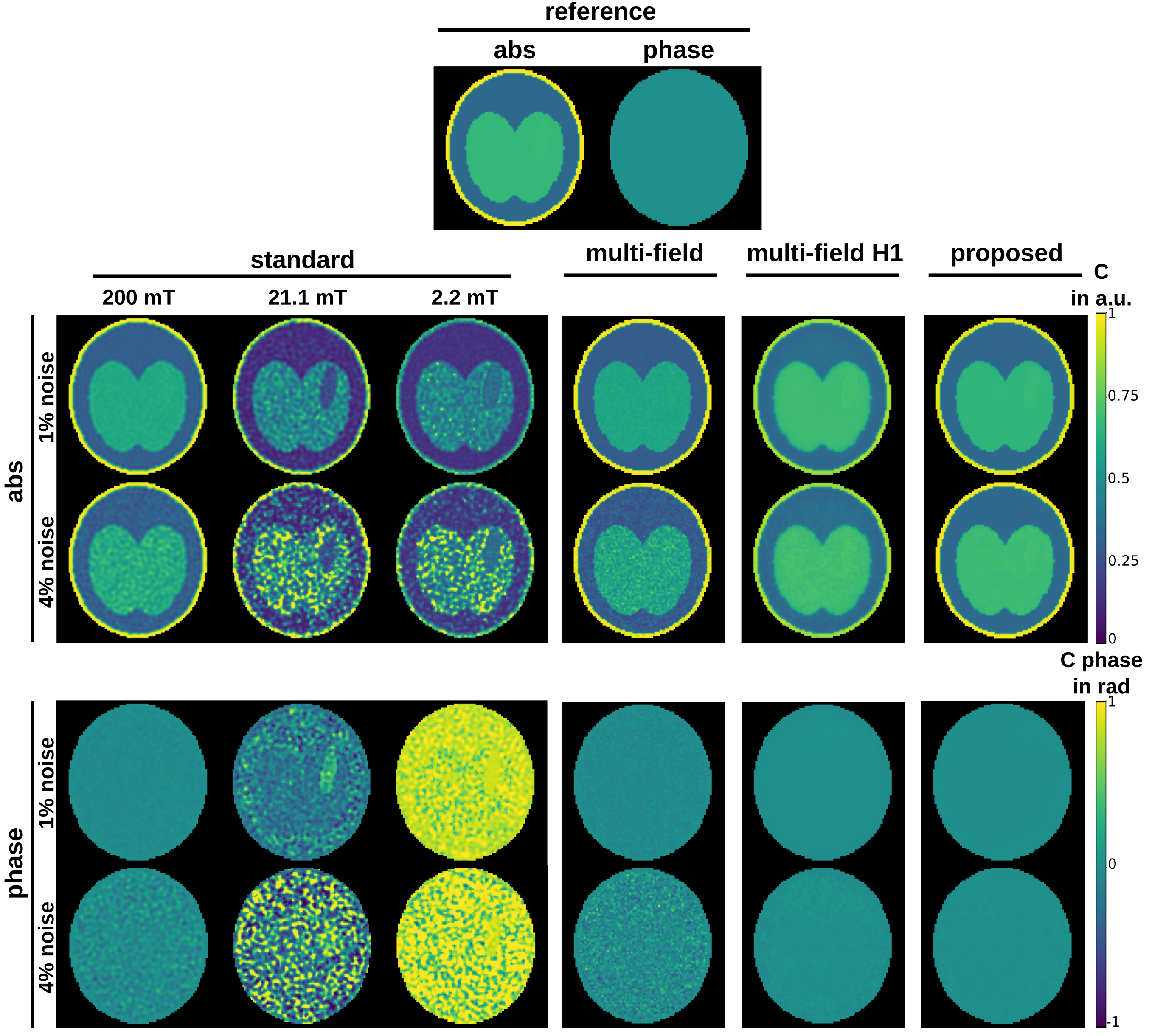} 
	\caption{Absolute value and phase of $C$ maps obtained from simulated FFC imaging inversion recovery data. The reference $C$ map is shown at the top. In the left column, the multi-field $C$ maps were obtained by pixel-wise fitting of each evolution field separately (standard), and the right column results from joint model-based reconstruction of all three evolution fields together (proposed). The noise level increases from top to bottom from 1~\% to 4~\%.}
	\label{fig:phantom_C_abs}
\end{suppfigure}

\begin{suppfigure}[ht]
	\centering
	\includegraphics[width=0.95\textwidth]{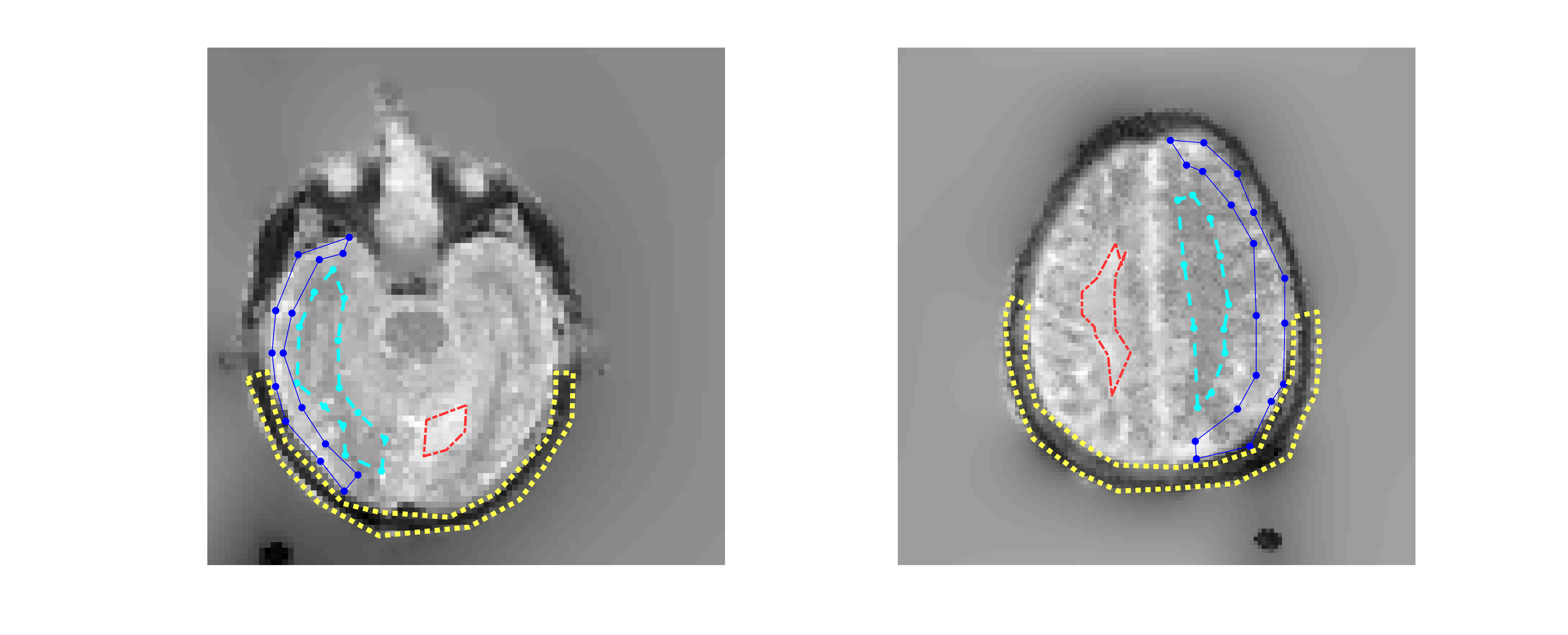} 
	\caption{Regions of interest selected to extract the dispersion profiles in Figure \ref{fig:invivo_dispersion_comparison} in patients I (left) and II (right). The regions for white matter are delineated in light blue dashed lines, grey matter in solid dark blue lines, fat in yellow dotted lines and lesions in red dot-dashed lines.}
	\label{fig:ROI_patients}
\end{suppfigure}

\begin{suppfigure}[ht]
	\centering
	\includegraphics[width=0.9\textwidth, height=0.85\textheight, keepaspectratio]{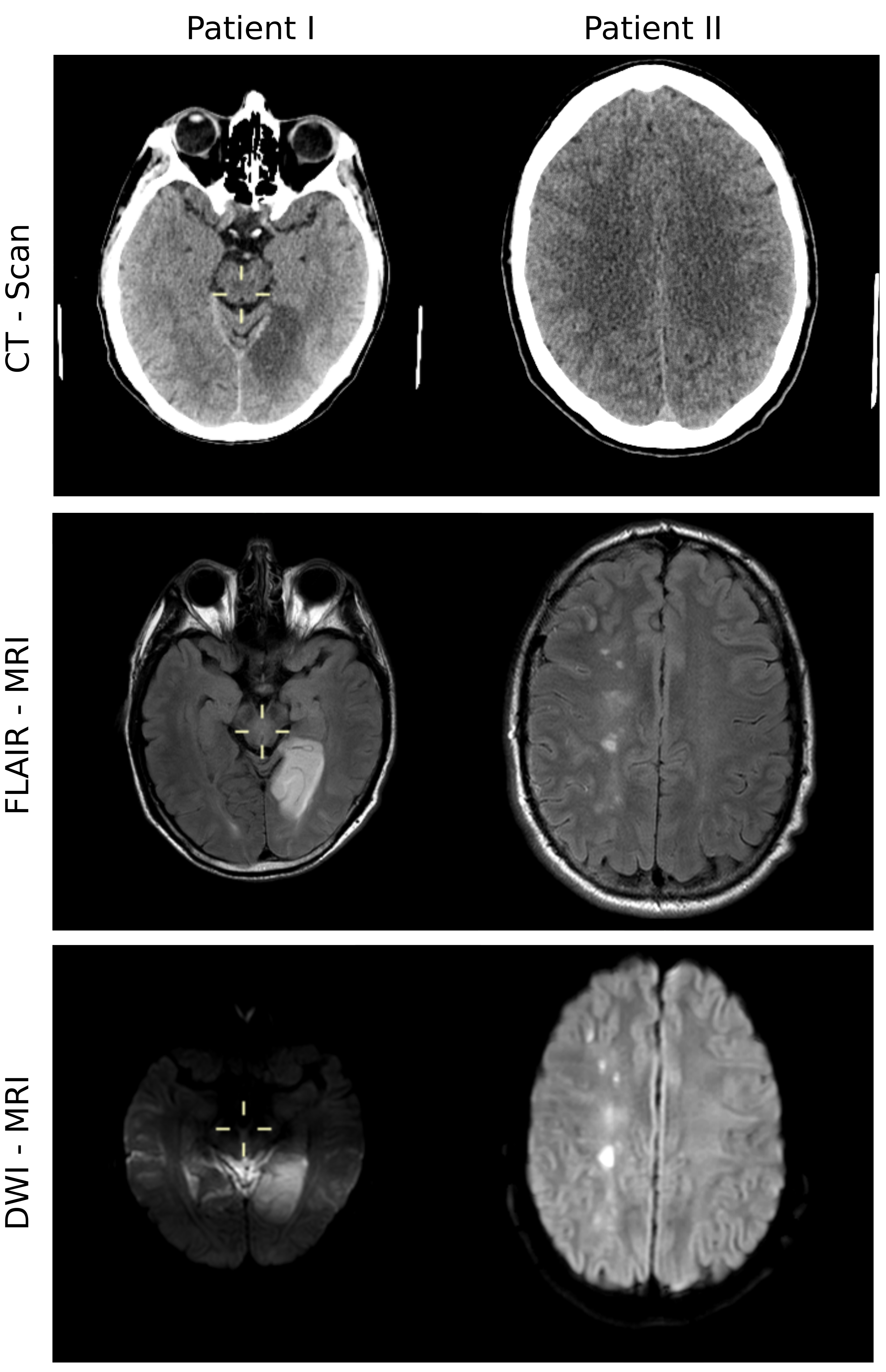}
	\caption{Images obtained from clinical examinations of Patient I (images on the left) and II (images on the right) from CT (top) and conventional MRI using 3D $T_2$-FLAIR (centre) and DWI (bottom), showing only the slice that corresponds with the single slice acquisition of the FFC acquisition. The $T_2$-FLAIR ($T_2$-weighted-Fluid-Attenuated Inversion Recovery) sequence had an isotropic resolution of 0.625 mm, a SPIR fat suppression, an inversion delay of 1650 ms, an echo time of 340 ms and a repetition time of 4800 ms. The DWI (Diffusion-Weighted Imaging) had an in-plane resolution of 0.8 mm, a slice thickness of 4 mm, a STIR fat suppression, an echo time of 77 ms, a repetition time of  3478 ms and a b-factor of 1000. The results indicate ischemic strokes and are clearly visible in both MRI FLAIR and DWI images, with patient II exhibiting multiple small strokes. CT scans are less informative for ischemic stroke, as illustrated here on patient II.}
	\label{fig:conventional}
\end{suppfigure}

\end{document}